\documentclass[printer]{aa}
\usepackage{txfonts}
\usepackage{deluxetable}
\usepackage{graphicx}
\usepackage{epstopdf}
\usepackage{auto-pst-pdf}
\immediate\write18{ls}
\DeclareGraphicsExtensions{.pdf,.png,.jpg,.ps,.eps}
\usepackage{graphics}
\usepackage{natbib}
\usepackage{float}
\usepackage{placeins}
\usepackage{multirow}
\usepackage{lscape}
\usepackage{tabularx}
\usepackage{longtable}
\bibpunct{(}{)}{;}{a}{}{,} 

\begin{document}
\newcommand{\msun}{\mbox{M$_{\odot}$}}
\newcommand{\rsun}{\mbox{R$_{\odot}$}}
\newcommand{\lsun}{\mbox{L$_{\odot}$}}
\newcommand{\tsun}{\mbox{T$_{\odot}$}}
\newcommand{\Inu}{\mbox{$I_{\nu}(b)$}}
\newcommand{\Snu}{\mbox{$S_{\nu}$}}
\newcommand{\Bnu}{\mbox{$B_{\nu}(\Td)$}}
\newcommand{\nInu}{\mbox{$I^{norm}(b)$}}
\newcommand{\nInum}{\mbox{$I^{norm}_{mod}(b)$}}
\newcommand{\kappanu}{\mbox{$\kappa_{\nu}$}}
\newcommand{\Tg}{\mbox{$T_{g}$}}
\newcommand{\Td}{\mbox{$T_{d}$}}
\newcommand{\Tk}{\mbox{$T_{K}$}}
\newcommand{\Tdr}{\mbox{$T_{d}(r)$}}
\newcommand{\Tkr}{\mbox{$T_{K}(r)$}}
\newcommand{\rhor}{\mbox{$\rho (r)$}}
\newcommand{\ppc}{pre-protostellar core}
\newcommand{\p}{\mbox{$p$}}
\newcommand{\nc}{\mbox{$n_c$}}
\newcommand{\chisq}{\mbox{$\chi_r^2$}}
\newcommand{\beam}{\mbox{$\theta_{mb}$}}
\newcommand{\router}{\mbox{$r_o$}}
\newcommand{\rinner}{\mbox{$r_i$}}
\newcommand{\rflat}{\mbox{$r_{flat}$}}
\newcommand{\msunmyr}{\mbox{M$_\odot$ Myr$^{-1}$}}
\newcommand{\ee}[1]{\mbox{${} \times 10^{#1}$}}
\newcommand{\eten}[1]{\mbox{$10^{#1}$}}

\newcommand{\h}{\mbox{$^h$}}
\newcommand{\m}{\mbox{$^m$}}
\newcommand{\s}{\mbox{$^s$}}
\newcommand{\degree}{\mbox{$^{\circ}$}}
\newcommand{\am}{\mbox{\arcmin}}
\newcommand{\as}{\mbox{\arcsec}}

\newcommand{\kms}{\mbox{km s$^{-1}$}}
\newcommand\cmv{\mbox{cm$^{-3}$}}
\newcommand\cmc{\mbox{cm$^{-2}$}}
\newcommand\cmdv{\mbox{cm$^{-2}$ (\kms)$^{-1}$}}
\newcommand{\um}{$\mu$m}

\newcommand{\x}{\mbox{${}\times{}$}}
\newcommand\tto{\mbox{$\rightarrow$}}
\newcommand\about{\mbox{$\sim$}}
\newcommand{\iras}{\mbox{\it IRAS}}
\newcommand{\iso}{\mbox{\it ISO}}
\newcommand{\ISO}{\mbox{\it ISO}}
\newcommand{\mm}{millimeter}
\newcommand\submm{submillimeter}
\newcommand\smm{submillimeter}
\newcommand\fir{far-infrared}
\newcommand\mir{mid-infrared}
\newcommand\nir{near-infrared}
\newcommand\uv{ultraviolet}
\newcommand{\sfr }{\mbox{$\dot M_{\star}$}}
\newcommand\sed{spectral energy distribution}
\newcommand{\ta}{{$T_A^*$}}
\newcommand{\tex}{\mbox{$T_{\rm ex}$}}
\newcommand{\tmb}{\mbox{$T_{\rm mb}$}}
\newcommand{\tr}{\mbox{$T_R$}}
\newcommand{\tk}{\mbox{$T_K$}}
\newcommand{\td}{\mbox{$T_d$}}
\newcommand{\lbol}{\mbox{$L_{bol}$}} 
\newcommand{\tbol}{\mbox{$T_{bol}$}} 
\newcommand{\dv}{\mbox{$\Delta v$}}
\newcommand{\n}{\mbox{$n$}}
\newcommand{\nbar}{\mbox{$\overline{n}$}}
\newcommand{\mv}{\mbox{$M_V$}} 
\newcommand{\mc}{\mbox{$M_N$}} 
\newcommand{\mn}{\mbox{$M_n$}} 
\newcommand{\meanl}{\mbox{$\langle l \rangle$}} 
\newcommand{\meandev}{\mbox{$\langle \delta \rangle$}} 
\newcommand{\meanar}{\mbox{$\langle a/b \rangle$}} 
\newcommand{\mean}[1]{\mbox{$\langle#1\rangle$}} 
\newcommand{\opacity}{\mbox{$\kappa(\nu)$}} 
\newcommand{\av}{\mbox{$A_V$}} 
\newcommand{\bperp}{\mbox{$B_{\perp}$}} 
\newcommand{\rinf}{\mbox{$r_{inf}$}} 
\newcommand{\fsmm}{\mbox{$L_{smm}/L_{bol}$}} 
\newcommand{\lsmm}{\mbox{$L_{smm}$}} 
\newcommand{\alphanir}{\mbox{$\alpha_{NIR}$}} 
\newcommand{\isrf}{\mbox{\rm{ISRF}}}
\newcommand{\sub}[1]{_{\rm #1}}

\newcommand{\hh}{\mbox{{\rm H}$_2$}}
\newcommand{\form}{H$_2$CO}
\newcommand{\water}{H$_2$O}
\newcommand{\ammonia}{\mbox{{\rm NH}$_3$}}
\newcommand{\nthp}{\mbox{{\rm N$_2$H}$^+$}}
\newcommand{\coo}{$^{13}$CO}
\newcommand{\cooo}{C$^{18}$O}
\newcommand{\coooo}{C$^{17}$O}
\newcommand{\hcop}{HCO$^+$}
\newcommand{\hcopi}{H$^{13}$CO$^+$}
\newcommand{\dcop}{DCO$^+$}


\title{Temperatures of dust and gas in S~140.\thanks{Based on Herschel observations. Herschel is an ESA space observatory with science instruments provided by European-led Principal Investigator consortia and with important participation from NASA.}}

\author{E. Koumpia\inst{\ref{inst1},\ref{inst2}}, P.M. Harvey\inst{\ref{inst8},\ref{inst2}}, V. Ossenkopf\inst{\ref{inst4}}, F.F.S. van der Tak\inst{\ref{inst1}}, B. Mookerjea\inst{\ref{inst5}}, A. Fuente\inst{\ref{inst6}}, C. Kramer\inst{\ref{inst7}}}

\institute{SRON Netherlands Institute for Space Research, Landleven 12, 
  9747 AD Groningen, The Netherlands; Kapteyn Institute, University of Groningen, 
  The Netherlands\label{inst1}
\and
Astronomy Department, University of Texas at Austin, 1 University Station C1400, Austin, TX 78712--0259, USA\label{inst8}
\and 
I. Physikalisches Institut der Universit\"{a}t zu K\"{o}ln, Z\"{u}lpicher Stra$\ss$e 77, 50937, K\"{o}ln, Germany\label{inst4}
\and 
Department of Astronomy and Astrophysics, Tata Institute of Fundamental Research, Homi Bhabha Road, Colaba, 400005, Mumbai, India\label{inst5}
\and 
Observatorio Astron\'{o}mico Nacional (OAN,IGN), Apdo 112, 28803, Alcal\'{a} de Henares, Spain\label{inst6}
\and 
Instituto Radioastronom\'{i}a Milim\'{e}trica, Av. Divina Pastora 7, Nucleo Central, 18012 Granada, Spain\label{inst7}
\and \email{e.koumpia@sron.nl;pmh@astro.as.utexas.edu}\label{inst2}
}
\date{Received date/Accepted date}

\begin{abstract}
{In dense parts of interstellar clouds ($\ge$ 10$^{5}$ cm$^{-3}$), dust and gas are expected to be in
thermal equilibrium, being coupled via collisions. However, previous
studies have shown that in the presence of intense radiation fields, the
temperatures of the dust and gas may remain decoupled even at higher
densities.} 
{The objective of this work is to study in detail the
temperatures of dust and gas in the photon-dominated region S~140,
especially around the deeply embedded infrared sources IRS~1--3 and at
the ionization front.} 
{We derive the dust temperature and column density by combining Herschel PACS continuum observations
with SOFIA observations at 37~$\mu$m and SCUBA data at 450~$\mu$m. 
We model these observations using simple greybody fits and the DUSTY radiative transfer code. 
For the gas analysis we use RADEX 
to model the CO 1--0, CO 2--1, $^{13}$CO 1--0 and C$^{18}$O 1--0 emission lines
mapped with the IRAM--30m telescope over a 4$'$ field. Around IRS 1--3, we use HIFI observations of
single-points and cuts in CO 9--8, $^{13}$CO 10--9 and C$^{18}$O 9--8
to constrain the amount of warm gas, using the best fitting dust model derived with DUSTY as input to the non--local radiative transfer model RATRAN. 
The velocity information in the lines allows us to separate the quiescent component
from outflows when deriving the gas temperature and column density.}
{We find that the gas temperature around the infrared sources varies between $\sim$35 and $\sim$55~K.
 In contrast to expectation, the gas is systematically warmer than the dust by $\sim$5--15~K despite the high gas density. 
In addition we observe an increase of the gas temperature from 30--35~K in the surrounding  up to 40--45~K towards the ionization front, most likely due to the UV radiation from the external star. Furthermore, detailed models of the temperature structure close to IRS~1 which take the known density gradient into account show that the gas is warmer and/or denser than what we model. Finally, modelling of the dust emission from the sub--mm peak SMM~1 constrains its luminosity to a few $\times$~10$^{2}$~$\lsun$.} 
{We conclude that the gas heating in the S~140 region is very efficient even at high densities. The most likely explanation is deep UV penetration from the embedded sources in a clumpy medium and/or oblique shocks.}

\end{abstract}

\keywords{ISM: individual (S~140), ISM: kinematics and dynamics, ISM: molecules, stars: formation}

\titlerunning{Temperatures of Dust and Gas in S~140.} 
\authorrunning{Koumpia et al} 
\maketitle


\section{Introduction} 

Stars are born inside dense molecular structures in the interstellar medium (ISM) which consist
of gas and dust. The dust grains in clouds associated with star--forming regions absorb the short--wavelength radiation from
the central stars, heat up, and then re--emit radiation at far--infrared (FIR) and
sub--millimeter wavelengths. The thermal
radiation from dust is optically thin at these wavelengths and thus is a good tracer of 
physical parameters such as temperature, density and gas mass of the clouds. Dust is
efficiently heated through near IR radiation due to the broadband absorption properties 
of the dust grains. On the other hand, the gas temperature T$_{\mathrm gas}$, is mainly governed 
by heating processes driven by UV radiation (photoelectric heating, H$_{2}$ excitation, dissociation). 

In the well--shielded centers of massive cores, 
the primary effects for the thermal balance are cosmic ray heating, molecular line cooling and collisional heating 
or cooling due to dust--gas collisions \citep{Goldsmith1978,Tielens2005,Draine2011}.

At densities above $\sim$10$^{4.5}$, the dust and gas are expected to be collisionally 
coupled and they are characterized by the same temperature \citep{Goldsmith01}. At lower densities, the rate of collisions 
between dust and gas decreases and the cooling of the gas via fine-structure and 
molecular rotational transitions becomes dominant (e.g. CO, $[$C II$]$ and $[$OI$]$) \citep{Sternberg95,Kaufman99,Meijerink05}.

Excitation of the rotational levels of molecules like CO (low dipole moment) observed at (sub)mm wavelengths is provided by collisions with H$_{2}$. If these collisions are frequent enough to exceed the spontaneous decay rate, the levels will get into thermal equilibrium with H$_{2}$. The critical densities n$_{cr}$ of CO and isotopologues for the J$=$1--0 and 2--1 transitions at 50~K are $\sim$2$\times$10$^{3}$~cm$^{-3}$ and $\sim$2$\times$10$^{4}$~cm$^{-3}$ respectively \footnote{The values were calculated using Einstein coefficients and collisional rates (Eq. (2)) from \citet{Yang10}.}.
At this point their excitation temperature will approach the gas kinetic temperature and thus the study of these lines is ideal for 
gas kinetic temperature estimates. The rarer isotopologues of CO, such as $^{13}$CO and C$^{18}$O, can be used in order to 
probe regions of high column density where the lines of the abundant isotopologues become optically thick. The critical densities n$_{cr}$ of CO and isotopologues for the J$=$9--8 and 10--9 transitions at 50~K are $\sim$1.2$\times$10$^{6}$~cm$^{-3}$ and $\sim$1.5$\times$10$^{6}$~cm$^{-3}$ respectively.

S~140 is a well studied HII region that lies at the south--west edge of the molecular cloud L~1204 
at a distance of 746~pc \citep{Hirota2008}. In previous studies \citep[e.g.][]{Dedes2010,Ikeda2011} a larger distance of 910~pc \citep{Crampton74} has been adopted. This edge is illuminated by the B0V star HD211880, creating a visible HII 
region and a Photon--Dominated Region (PDR) at the ionization front (IF). The cloud also hosts a cluster 
of embedded high--mass young stellar objects (YSOs), IRS~1 to 3 \citep{Evans1989} at a projected distance of $\sim 75$\arcsec~northeast of the IF. A large number of low--mass stars are also forming, making S~140 an ideal laboratory 
for studying star formation of various masses and the differences caused by irradiation 
from external (B0V star) and internal sources (IRS~1 to 3). 

The analysis of physical models around IRS~1 has been performed by a number of
previous authors including \citet{Harvey78}, \citet{Guertler91}, \citet{Minchin93}, \citet{vanderTak00}, \citet{Mueller02}, \citet{deWit09}, \citet{Maud13}. \citet{Poelman06} have reported the clumpiness towards S~140 giving a value of n$_{H_2}$ $\sim$ 10$^{4}$ cm$^{-3}$ for the interclump medium and~ $\sim$4$\times$10$^{5}$ for the clump gas.
Previous models lacked spatial information at many wavelengths and many of them were based on the assumption that the gas temperature is coupled to the dust as a result of the high densities \citep[e.g.][]{Poelman06}. 

In this study we have analyzed the temperature and column density around the S~140 high--mass star--forming region of embedded young stellar objects by combining Herschel PACS and HIFI data with ground--based mapping observations of mm--wave lines using the IRAM 30~m telescope. The new PACS continuum data and IRAM spectroscopic maps provide us with the 
highest angular resolution spatial information available covering an area that includes the YSOs in both datasets. The IF was covered by our larger IRAM maps (4\arcmin$\times$4\arcmin) but not by the PACS observations, since the latter covers a smaller area (45\arcsec$\times$45\arcsec). Furthermore, the data give us the opportunity to study independently the gas and dust temperatures around FIRS 1--3 and check whether the assumption of well coupled dust--gas is valid throughout this cloud.

In the following sections, we present the dust and gas modeling, and the resulting parameters. We present the observations in \S 2, the observational results in \S 3, the gas -- dust analysis and the comparison between gas and dust in \S 4, the effect of density gradients using more advanced dust and gas models in \S 5 and a discussion of our results and conclusions in \S 6.


\section{Observations and data reduction} 


\subsection{PACS data}

To investigate the temperature and column-density structure of the dust we have analyzed a single footprint of PACS \citep{Poglitsch10} 
with its 5$\times$5 spatial array of $\sim 9$\arcsec~pixels over a range
of wavelengths from $\sim 70~ \mu$m to $\sim 200~ \mu$m (obsid: 1342222256). The data reduction was performed using HIPE 9.1. 

Since the PACS spectrometer continuum data cover only the peak of the spectral energy distribution (SED) 
for S~140, we have supplemented our analysis with shorter and
longer wavelength images. We used the SOFIA/FORCAST images at 11.1, 31.5, and 37
$\mu$m discussed by \citet{Harvey12} and the 24.5 $\mu$m Subaru/COMICS data of \citet{deWit09}. 
Since all of these observations were obtained with resolution better than
that of the PACS/Spec images, we used them both at their native resolution and re-convolved to the PACS/Spec 187 $\mu$m resolution. 
These raw and convolved images were
also re-sampled to a 1\arcsec~grid like the PACS/Spec images. Longward of the PACS wavelengths 
the highest resolution image available is that in the JCMT SCUBA archive\footnote{m96bu47199708210025 from http://www.cadc-ccda.hia-iha.nrc-cnrc.gc.ca/en/jcmt/} at 450~$\mu$m. We have used this image
without further processing in our analysis below. We have also examined the SCUBA 850~$\mu$m image which is qualitatively similar but with lower spatial resolution.

\subsection{IRAM data}

The 30m observations were conducted during 4 hours on September 24th,
2011. The Eight MIxer Receiver (EMIR) was used during the observations. CO $J$ = 1 $\rightarrow$ 0 and isotopologues ($\sim$ 110--115 GHz, 3~mm) and CO $J$ = 2 $\rightarrow$ 1 (230.54 GHz, 1~mm) map data 
were gathered in two linear polarizations. 
Our 4\arcmin$\times$4\arcmin~maps were centered close to S~140-IRS~1 position at RA = 22:19:18.30, Dec = 63:18:54.2 (J2000) (Figure~\ref{fig:spectrum7.ps}). 
For this work, we selected the maps of CO $J$ = 1 $\rightarrow$ 0 (115.27 GHz),
$J$ = 2 $\rightarrow$ 1 (230.54 GHz) together with the isotopologues $^{13}$CO $J$ = 1 $\rightarrow$ 0 (110.20 GHz) and C$^{18}$O $J$ = 1 $\rightarrow$ 0 (109.78 GHz). At all the observed lines, the FTS and WILMA were connected in parallel as backends.  
In the 3~mm set up we collected data from 3 bands of $\sim 4$ GHz each including the isotopic lines of $^{13}$CO $J$ = 1 $\rightarrow 0$ and C$^{18}$O $J$ = 1 $\rightarrow$ 0 in the CO setup. The beam sizes are $\sim 21$\arcsec~at 3~mm and $\sim 11$\arcsec~at 1~mm. 
 
The
sky opacity measured by the taumeter at 225\,GHz was stable at
$\tau_{225}=-0.26$.
Pointing and focus were checked about every hour on quasars and on
Uranus, and were
stable within $2"$ and $0.2\,$mm. The spectral line calibration was
checked by pointed
observations on IRS~1 and IF. The on--the--fly mapping observations were
done in perpendicular
directions to avoid scanning artefacts.
The typical system temperatures were 96--215~K at 3~mm, and 175 -- 190~K
at 1~mm. The resulting rms falls between 0.152--0.705~K at 3~mm and $\sim$ 0.525~K at 1~mm for a
velocity resolution of $\sim$ 0.5~km~s$^{-1}$ and $\sim$0.25~km~s$^{-1}$ respectively.

The intensities were converted from antenna temperature units to a scale of main-beam temperature (T$_{mb}$), 
dividing by the main-beam efficiencies $\eta_{mb}$ (B$_{eff}$/F$_{eff}$) of 0.82 and 0.64 at 3~mm and 1~mm respectively. 
Velocities are given with respect to the LSR. Baselines of orders 0 up to 4 have been subtracted. Data reduction and analysis were performed using the GILDAS
software \footnote{GILDAS is a radio-astronomy software developed by IRAM. See http://www.iram.fr/IRAMFR/GILDAS/}.

\subsection{HIFI data}

In order to test the reliability of our method for the gas analysis, we use single pointing observations (spectra) (Figure~\ref{fig:overplot_hifi.ps}) and cuts from Herschel--HIFI (Figure~\ref{fig:spectrum7.ps}, top), in addition to the 4'$\times$4' maps from IRAM--30m. 
Single-pointing, frequency--switching spectra in bands 1a, 1b, 2b, 3a, 4a, 4b, 6b and 7b and
OTF maps were analyzed. The single pointing data were taken towards
S~140-IRS~1 at RA = 22:19:18.21, Dec = 63:18:46.9 (J2000)
and towards a position in the IF at RA = 22:19:11.53,
Dec = 63:17:46.9 (J2000). 

For this analysis, we used only the observations of CO $J$ = 9 $\rightarrow$ 8 (1036.9 GHz), together with the isotopologues, $^{13}$CO $J$ = 10 $\rightarrow$ 9 (1101.35 GHz), C$^{18}$O $J$ = 9 $\rightarrow$ 8 (987.56 GHz), towards IRS~1 and IF (obsids: 1342195050, 1342196426, 1342219209, 1342195049, 1342201741). In addition we used the cuts of CO 9$\rightarrow$8 and $^{13}$CO $J$ = 10 $\rightarrow$ 9 (obsids:1342201741, 1342201806).
The observed line parameters in units of antenna temperatures, have been corrected in units of main beam temperatures using the main beam efficiency of 0.74 (987.56 GHz, 1036.9 GHz, 1101.35 GHz) \citep{Roelfsema2012}, while the beam size at these frequencies is $\sim 21$\arcsec.






\section{Observational results} 

\subsection{Dust continuum observations}

\begin{figure}[ht]
\begin{center}
\includegraphics[width=2.5in]{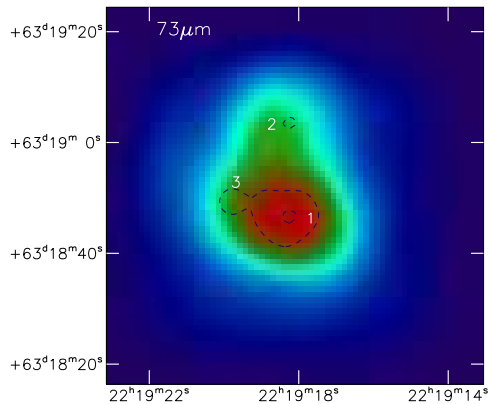} \\
\end{center}
\caption{PACS/Spec continuum image at 73 \um: 5$\times$5 spatial array of
9\arcsec~pixels re--sampled to 1\arcsec~grid with
contours of 37$\mu$m (SOFIA) emission overlaid showing the positions of IRS~1, 2, and 3.}
\label{fig:1.eps}
\end{figure}

For this analysis we selected our 45\arcsec$\times$45\arcsec maps in a number of continuum wavelengths with no obvious 
line emission or with line emission that was easily masked. The particular wavelengths chosen were centered on the wavelengths for which
convolution kernels have been developed by \citet{Aniano11}, which enables us to
convolve all maps to a common resolution. These wavelengths PACS/Spec data
are 73, 75, 84, 94, 110, 125, 136, 145, 150, 168, and 187~ $\mu$m. We extracted 3~$\%$ bandwidth
continuum fluxes centered on each of these wavelengths while masking out obvious line
emission. For comparison with other wavelength data
as well as for comparison with modeling results, we have re-sampled all these images to a 1\arcsec
spatial grid (Figure~\ref{fig:1.eps}). For the final comparison the model images are convolved to a telescope PSF
and typically a 9\arcsec~square pixel to simulate the PACS/Spec observations.

Figure~\ref{fig:1.eps} shows the image at 73\um. The 37$\mu$m (SOFIA) emission is overlaid using black contours showing the positions of IRS~1, 2, and 3. As shown in the image, the emission from IRS~1 dominates at this wavelength. 

\subsection{The spatial and velocity distribution of CO emission}


\begin{figure}[ht]
\begin{center}$
\begin{array}{c}
\includegraphics[width=2.55in]{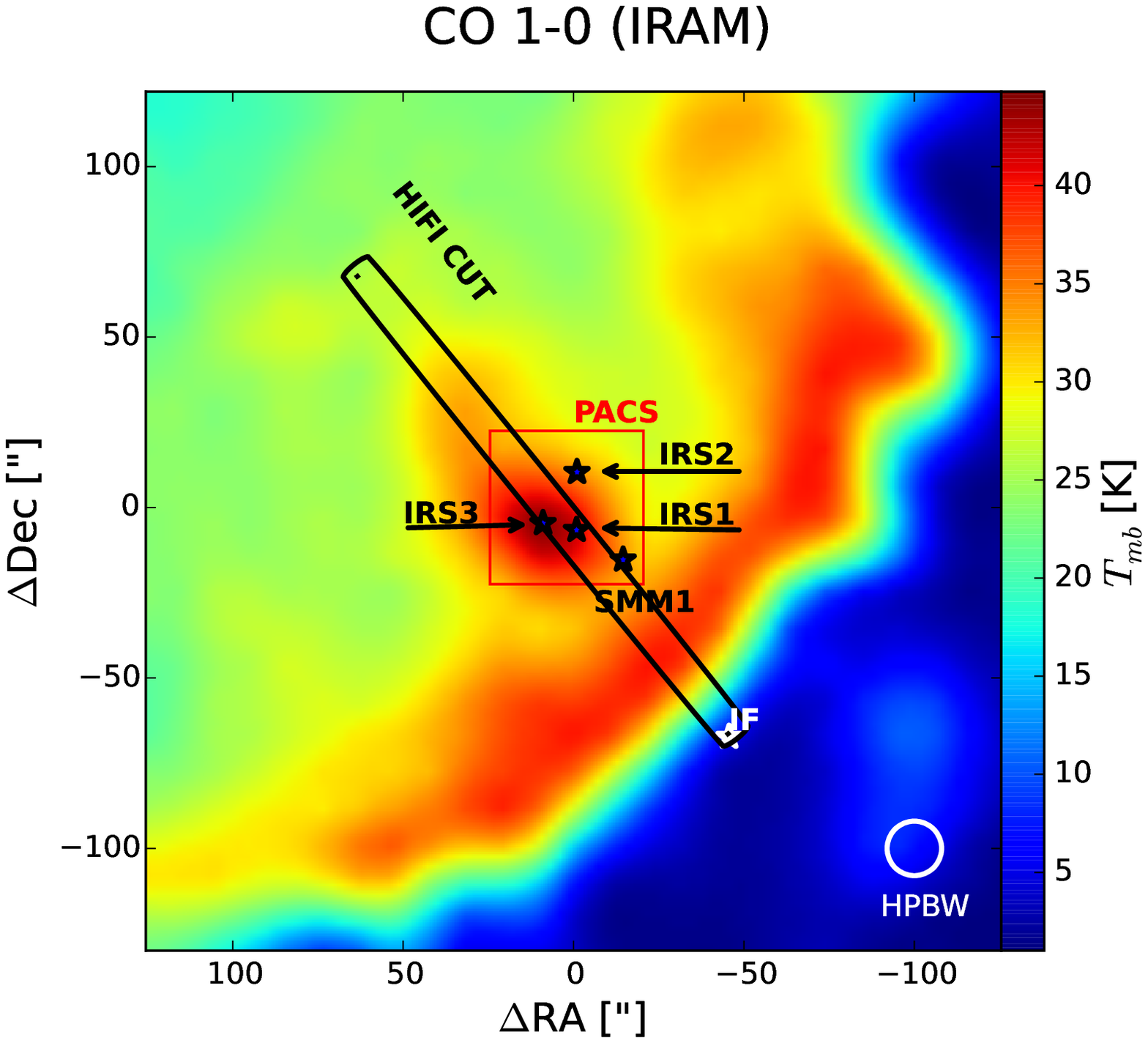} \\
\includegraphics[width=2.55in]{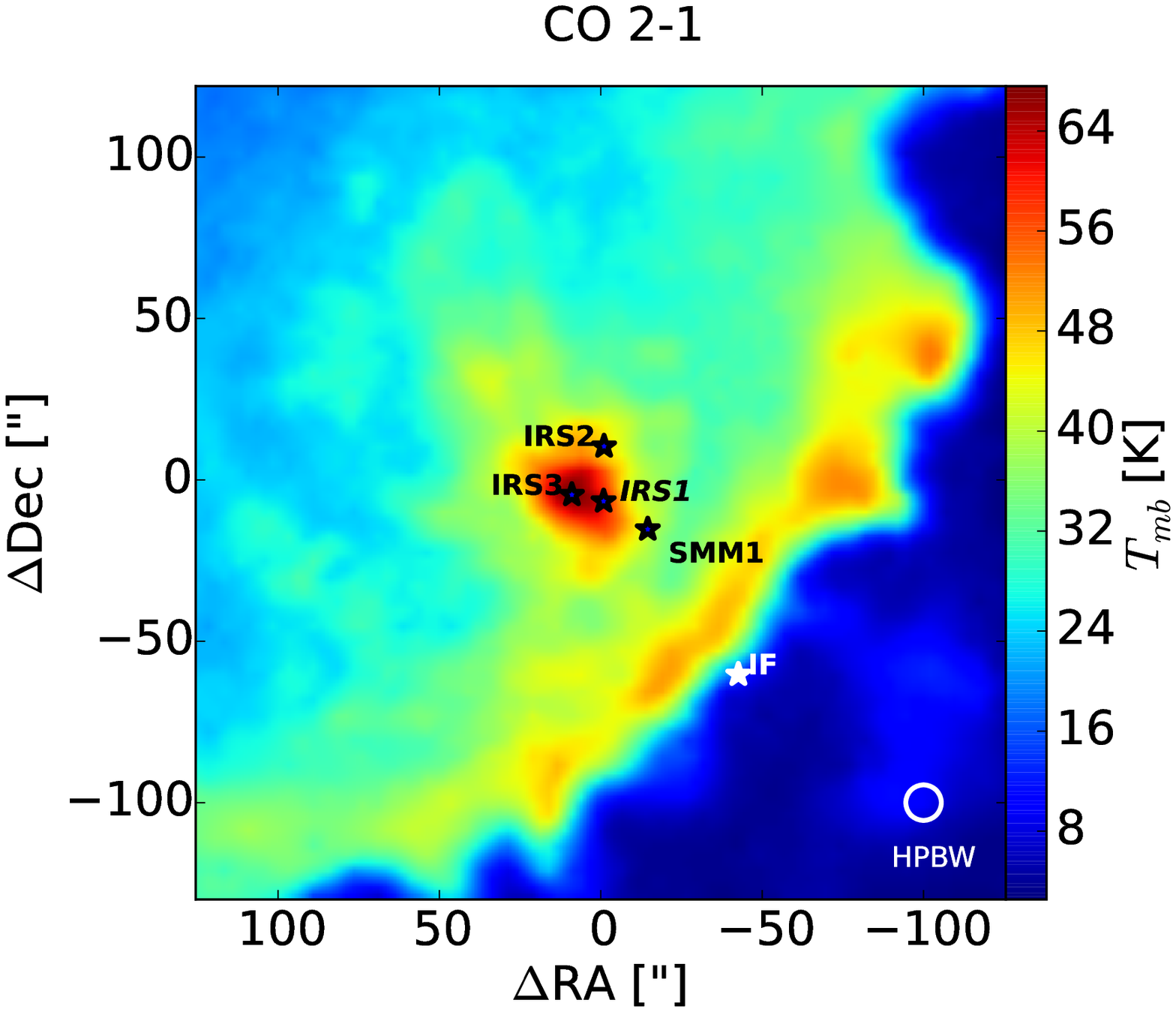} \\ 
\includegraphics[width=2.55in]{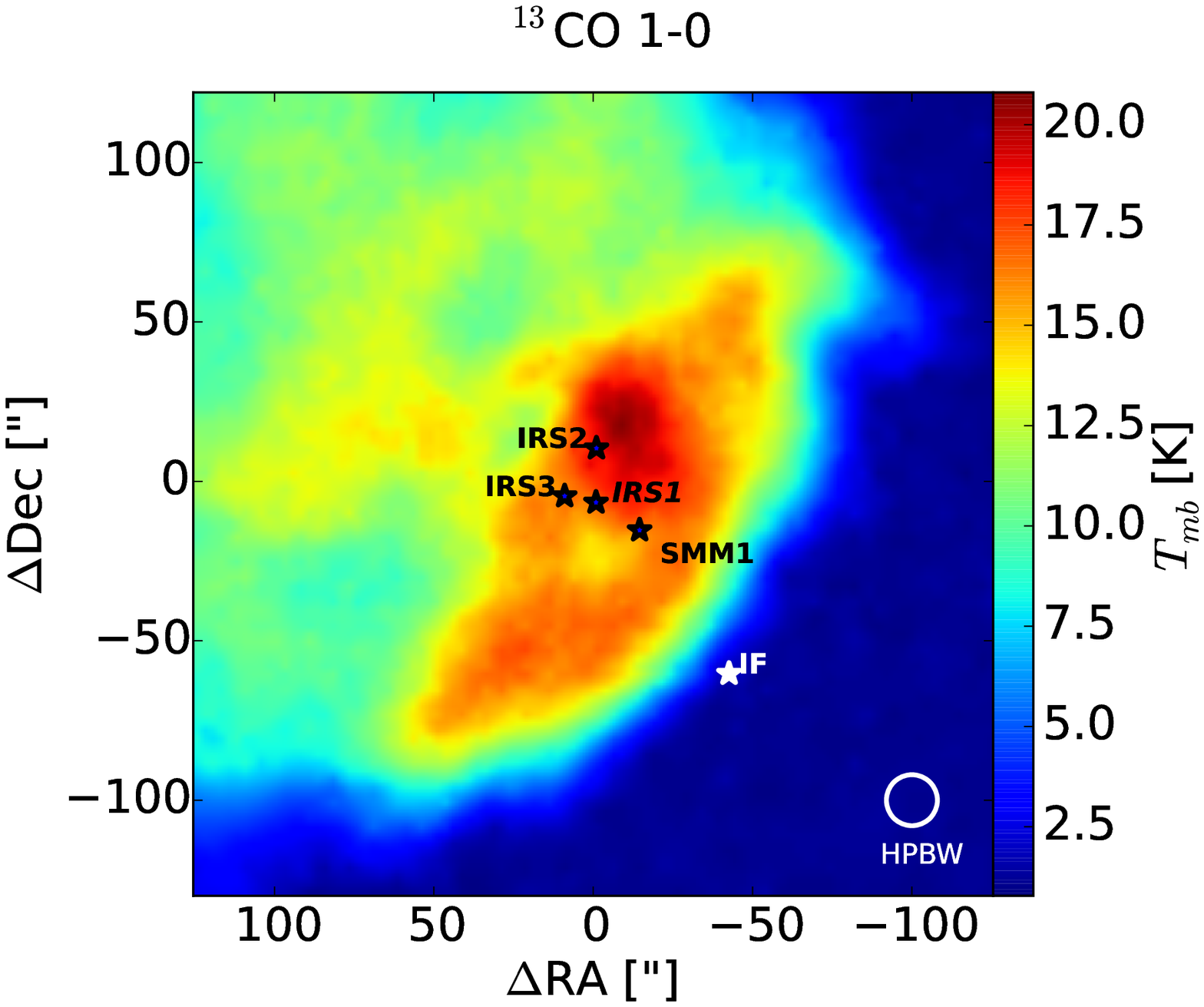} \\
\includegraphics[width=2.55in]{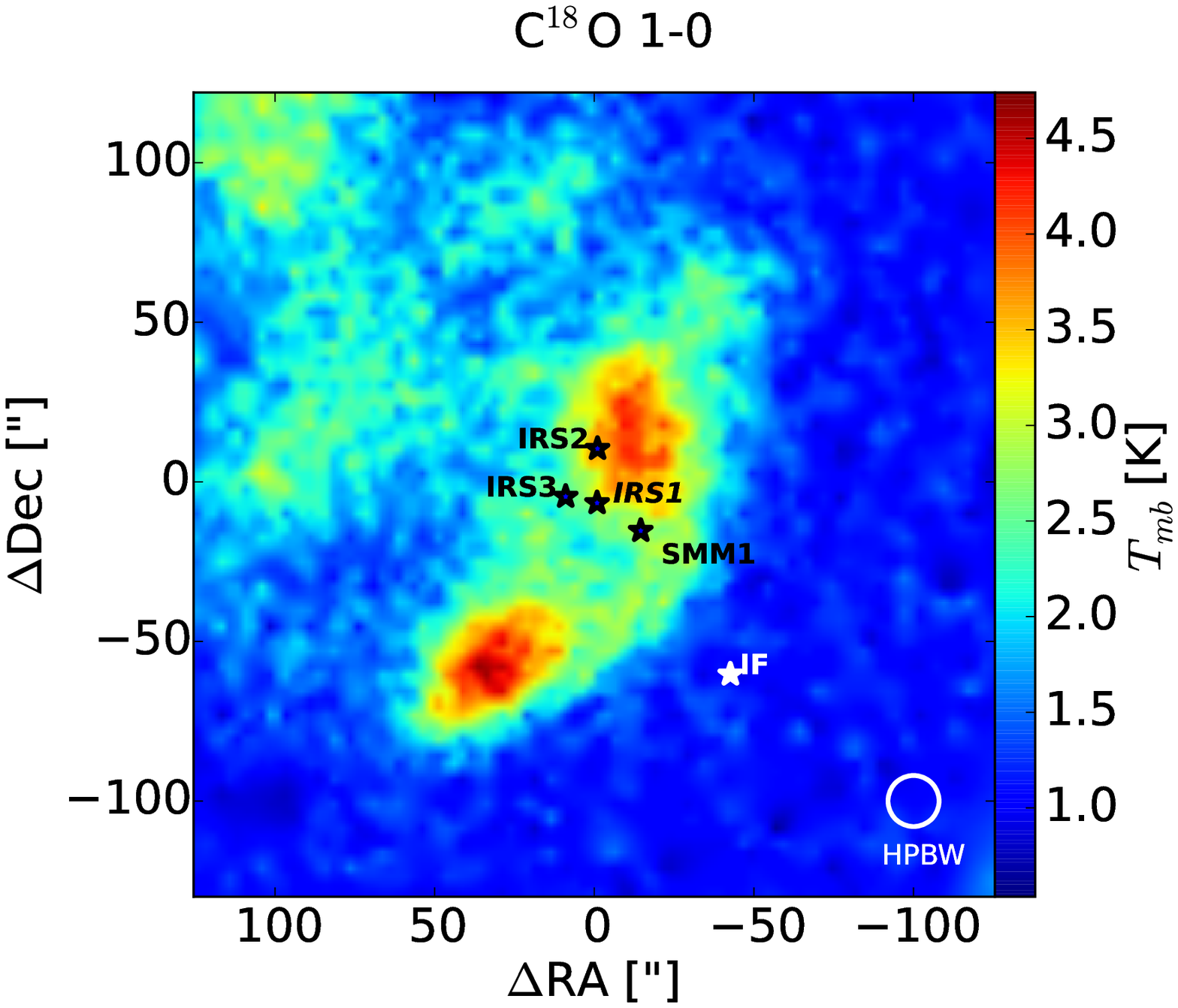}
\end{array}$
\end{center}
\caption{Spatial distribution of CO 1--0, CO 2--1, $^{13}$CO 1--0 \& C$^{18}$O 1--0 peak intensities (HIFI cut \& PACS overplotted). The IRS~1--3 positions were taken from \citet{Evans1989} \& SMM~1 from \citet{Maud2013}. The center is at RA $=$ 22:19:18.30, Dec $=$ 63:18:54.2 (J2000).}
\label{fig:spectrum7.ps}
\end{figure}

\begin{figure}[ht]
\includegraphics[scale=2.0]{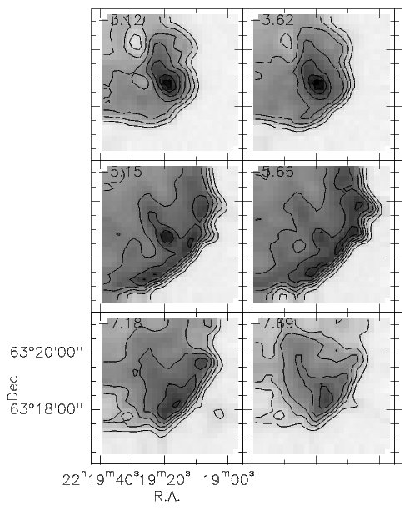}
\caption{Channel maps of CO 1--0 obtained with IRAM, with the central velocities given in km$\;$ s$^{-1}$. The levels of contours are 45~K (darker) till 10~K (lighter) using a step of 5~K.}
\label{fig:spectrum6.ps}
\end{figure}


Figure~\ref{fig:spectrum7.ps} shows the maps of the peak intensities as observed in their original angular resolution at 
their peak velocities of 
CO 1--0, CO 2--1, $^{13}$CO 1--0 and C$^{18}$O 1--0. The lines of the
main isotope show two clearly disjunct peaks, one at the densest cluster
between the YSOs IRS~1 and IRS~3, and one marking the external interface of
the cloud illuminated by HD211880. As $^{13}$CO and C$^{18}$O trace the column density structure, the interface peak is shifted
into the cloud for these lines, merging with the peak around
the embedded cluster. 

Finally, while CO 1--0 and CO 2--1 peak towards IRS~1--3 ($\Delta$RA: 10.70\arcsec, $\Delta$Dec: -4.0\arcsec), the $^{13}$CO 1--0 and C$^{18}$O 1--0 peak closer to IRS~2 ($\Delta$RA: -11.20\arcsec, $\Delta$Dec: 18.30\arcsec).
C$^{18}$O 1--0 shows a second peak southeast of IRS~1. This reflects a column density effect which is also revealed later in the column density map (see Figure~\ref{fig:column_density_map_final.ps}).

Figure~\ref{fig:spectrum6.ps} shows channel maps of CO 1--0 to
illustrate the velocity structure. We find a clear velocity offset
between the gas around the cluster and the interface and a much narrower
velocity distribution of the interface gas. Figures~\ref{fig:overplot.ps}--\ref{fig:overplot1.ps} shows 
the line profiles of CO and isotopologues as observed towards IRS~1 and IF. The peak velocity of the emission changes with position in the map. 
The IF peaks close to the velocity of the source which is $\sim -7.1$~km$\;$ s$^{-1}$, while IRS~1 peaks at $\sim -3.6$~km$\;$ s$^{-1}$.
The difference may be caused by outflows driven by the infrared sources. For the above reasons we chose to model the peak intensities of the lines in their peak velocities.


\subsection{Line profiles} 

\begin{figure}[ht]
\includegraphics[scale=0.33, angle=90]{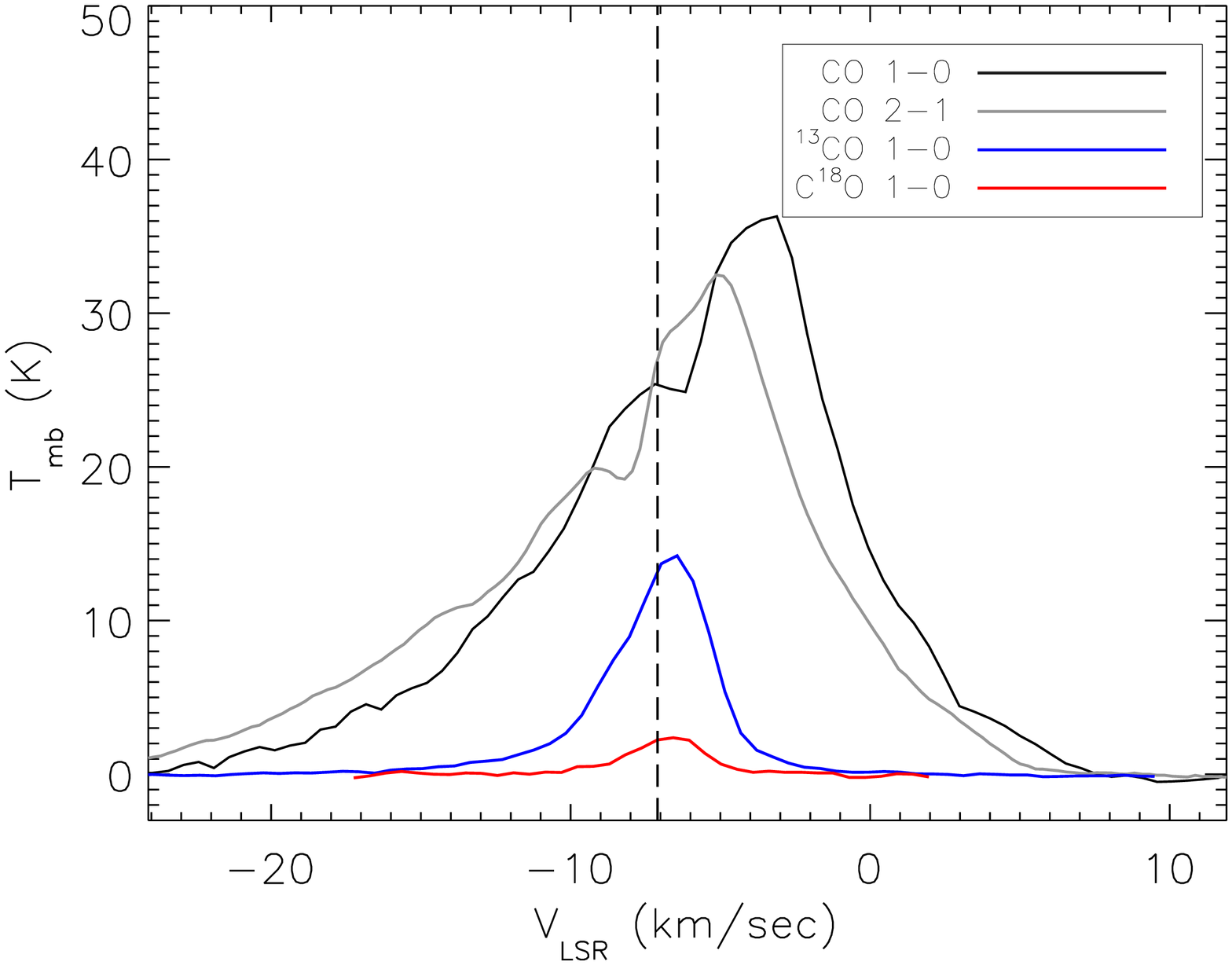}
\caption{Line profiles of CO 1--0 (black line), CO 2--1 (grey line), $^{13}$CO 1--0 (blue line) and C$^{18}$O 1--0 (red line) towards IRS~1 as observed with IRAM (offset position: 0\arcsec, -4.0\arcsec). Wings are visible at blueshifted velocities. The dashed vertical line represent the velocity of the source (-7.1~km$\;$ s$^{-1}$).}
\label{fig:overplot.ps}
\end{figure}

\begin{figure}[ht]
\includegraphics[scale=0.33, angle=90]{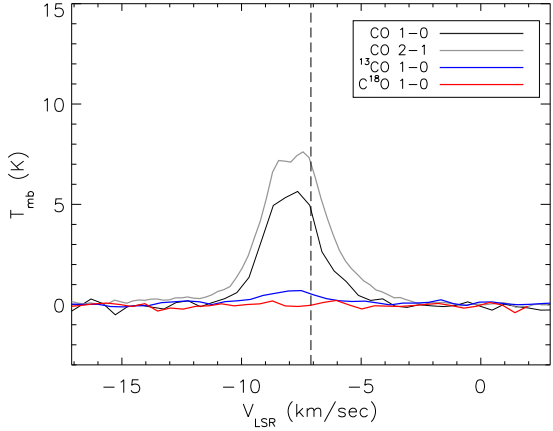}
\caption{Line profiles of CO 1--0 (black line), CO 2--1 (grey line), $^{13}$CO 1--0 (blue line) and C$^{18}$O 1--0 (red line) towards the IF as observed with IRAM (offset position: -42.7\arcsec, -68.0\arcsec). The dashed vertical line represent the velocity of the source (-7.1~km$\;$ s$^{-1}$).}
\label{fig:overplot1.ps}
\end{figure}

\begin{figure}[ht]
\includegraphics[scale=0.33, angle=90]{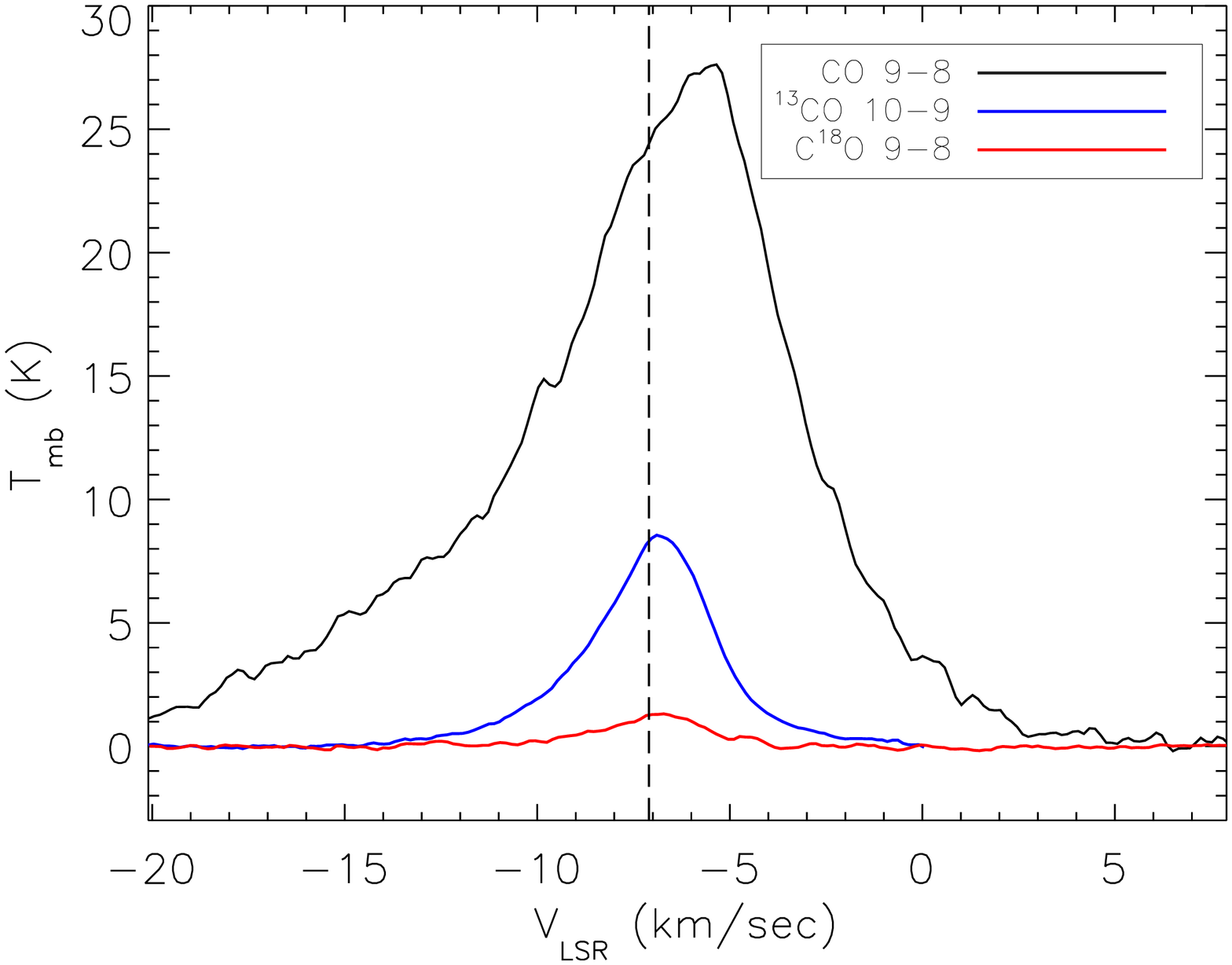}
\caption{Line profiles of CO 9--8 (black line), $^{13}$CO 10--9 (blue line) and C$^{18}$O 9--8 (red line) towards IRS~1 as observed with HIFI (RA = 22:19:18.21, Dec = 63:18:46.9). The dashed vertical line represent the velocity of the source (-7.1~km$\;$ s$^{-1}$).}
\label{fig:overplot_hifi.ps}
\end{figure}

The line profiles as shown in Figures~\ref{fig:overplot.ps} and \ref{fig:overplot1.ps}, are stronger towards IRS~1 and weaker towards the IF (Table~\ref{co_lines}). 
At the individual positions, the lines from CO are broader than the lines from the other isotopologues of CO (Figure~\ref{fig:overplot.ps} and Figure~\ref{fig:overplot_hifi.ps}). 


\begin{figure}[ht]
\includegraphics[scale=0.35, angle = 270]{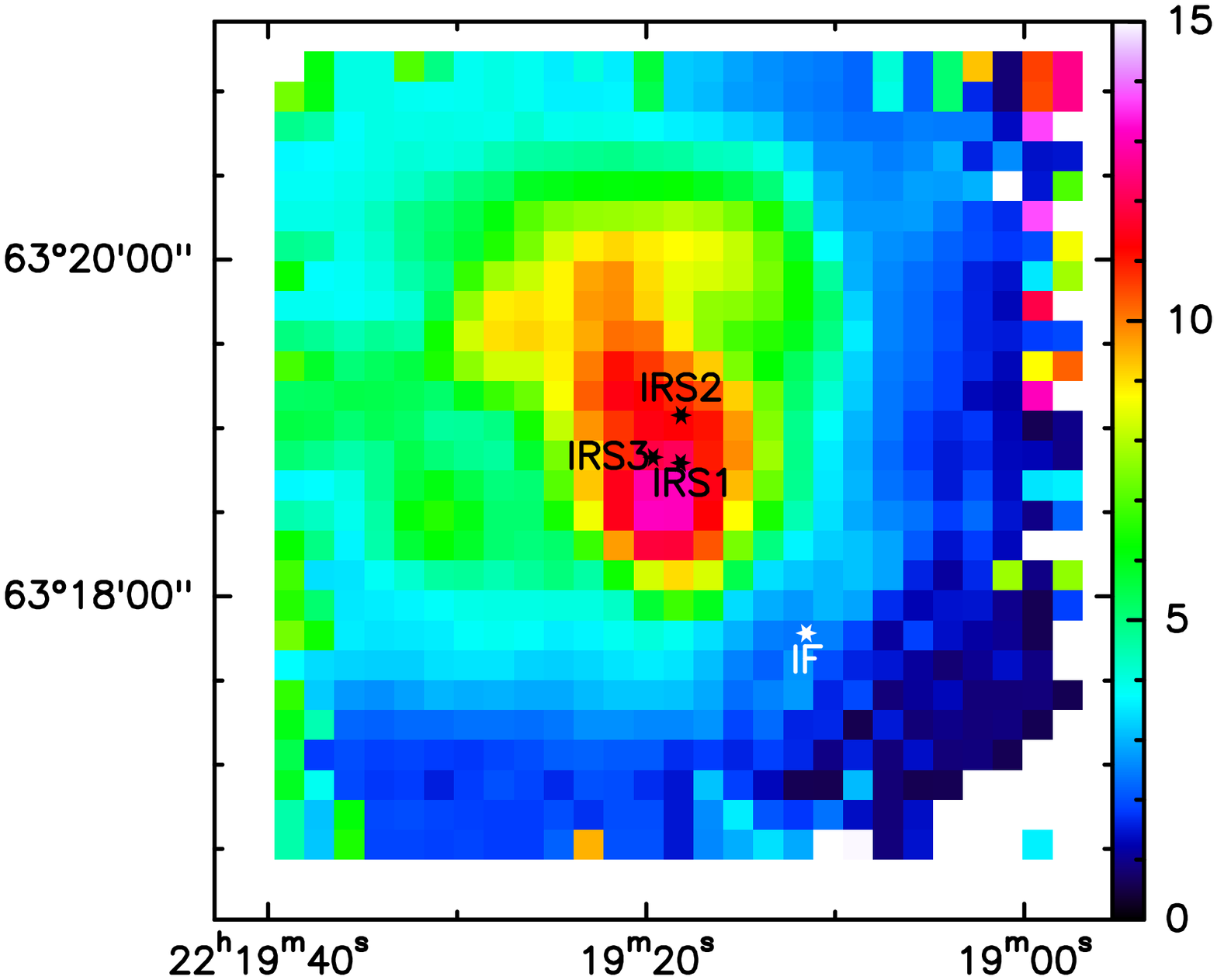}
\caption{Line width as observed throughout the cloud. The FWHMs were derived
from a 2nd moment map of the line emission. Broader profiles appear 
towards the center ($>$10~km$\;$s$^{-1}$), where the three infrared sources are located and narrower towards the ionization front ($<$5~km$\;$s$^{-1}$). The FWHM that characterizes the gas that surrounds the protostars varies between $\sim$5--10~km$\;$s$^{-1}$.}
\label{fig:FWHM}
\end{figure}

Table~\ref{co_lines} presents the line parameters as measured towards the IRS~1 and IF positions, applying a single component Gaussian fit, including both HIFI (Figure~\ref{fig:overplot_hifi.ps}) and IRAM (Figures~\ref{fig:overplot.ps} and \ref{fig:overplot1.ps}) observations. The uncertainties quoted for the peak intensities correspond to the observational RMS and the uncertainties for the FWHM and $\upsilon_{\rm LSR}$ are from the Gaussian fits to the peak position.
The width of the lines varies throughout the cloud, showing broader profiles 
towards the center of the map, where the three infrared sources are located and narrow towards the 
ionization front (Figure~\ref{fig:FWHM}). 
Being interested in the quiescent gas and thus the narrow component of the lines, 
we use observed and computed peak intensities in our calculations, as the narrow component dominates the total peak intensity of the lines ($\geq$50\%). With this method we limit the effects from outflow activities on our calculations but we cannot totally exclude them. The outflow contamination of the line is stronger towards the central sources where the integrated contribution of the broad and narrow components is comparable and weaker in positions away from the sources where the narrow component provides 70--85\% of the total peak intensity. We
restricted the peak intensity analysis to a velocity window of
the width of the FWHM of the lines around the $^{13}$CO 1--0 line. In this way we always model the same quiescent component.


\begin{table*}[t]
\caption{Line parameters of CO and isotopologues towards the IRS~1 and IF positions -- IRAM \& HIFI data.}
\small
\centering
\setlength\tabcolsep{2pt}
\begin{tabular}{c c c c| c c c| c c c c c c c}
\hline
 & & & & & \bf IRS~1 & & & \bf IF (-45$\arcsec$,-60$\arcsec$) & \\
\hline\hline
Molecule & Transition & Rest Frequency & E$_{up}$ & $\upsilon_{\rm LSR}$ & FWHM & T$_{mb}$ & $\upsilon_{\rm LSR}$ & FWHM & T$_{mb}$ \\  &  & (MHz) & (K) & (km~s$^{-1}$) & (km~s$^{-1}$) & (K) & (km~s$^{-1}$) & (km~s$^{-1}$) & (K)\\
\hline\hline
CO & 1--0 & 115271.20 & 5.53 & $-5.07\pm0.08$ & $10.82\pm0.20$ & $31.91\pm0.55$ & $-7.32\pm0.19$ & $2.66\pm0.07$ & $5.80\pm0.08$\\
CO & 2--1 & 230538.00 & 16.60 & $-5.0\pm0.1$ & $11.40\pm0.22$ & $31.69\pm0.65$ & $-7.77\pm0.01$ & $3.20\pm0.02$ & $7.62\pm0.06$\\
CO & 9--8 & 1036912.39 & 248.88 & $-6.60\pm0.02$ & $8.32\pm0.07$ & $27.62\pm0.37$ & $-$7.80$\pm0.07$ & 2.50$\pm0.20$ & 1.20$\pm0.20$ \\
$^{13}$CO & 1--0 & 110201.35 & 5.29 & $-6.87\pm0.03$ & $4.09\pm0.08$ & $16.52\pm0.34$ &$-8.12\pm0.37$ & $1.34\pm0.70$ & $0.84\pm0.26$ \\
$^{13}$CO & 10--9 & 1101349.60 & 290.79 & $-6.98\pm0.03$ & $3.80\pm0.07$ & $8.56\pm0.03$ & \ldots & \ldots & $<$3~RMS\\
C$^{18}$O & 1--0 & 109782.17 & 5.27 &  $-5.71\pm0.14$ & $2.37\pm0.39$ & $2.88\pm0.37$ & \ldots & \ldots & $<$3~RMS\\
C$^{18}$O & 9--8 & 987560.20 & 237.02 & $-6.94\pm0.06$ & $3.26\pm0.04$ & $1.31\pm0.06$ & \ldots & \ldots & $<$3~RMS\\
\hline\hline
\label{co_lines}
\end{tabular}
\label{co_lines}
\end{table*}


\section{Physical conditions} 

\subsection{RADEX fitting} 

\subsubsection{Method}

We use the non--LTE radiative transfer program RADEX \citep{vanderTak07} 
to compare the observed line intensity ratios with a grid of models for deriving
kinetic temperatures, gas densities and column densities.
As model input we use the molecular data from 
the LAMDA database \citep{Schoier05} and CO collisional rate coefficients from \citet{Yang10}.
RADEX predicts line intensities of a molecule for a given set of parameters: kinetic temperature, column density, H$_{2}$ density, background temperature and line width.



\begin{figure}[h]
\includegraphics[scale=0.29, angle=90]{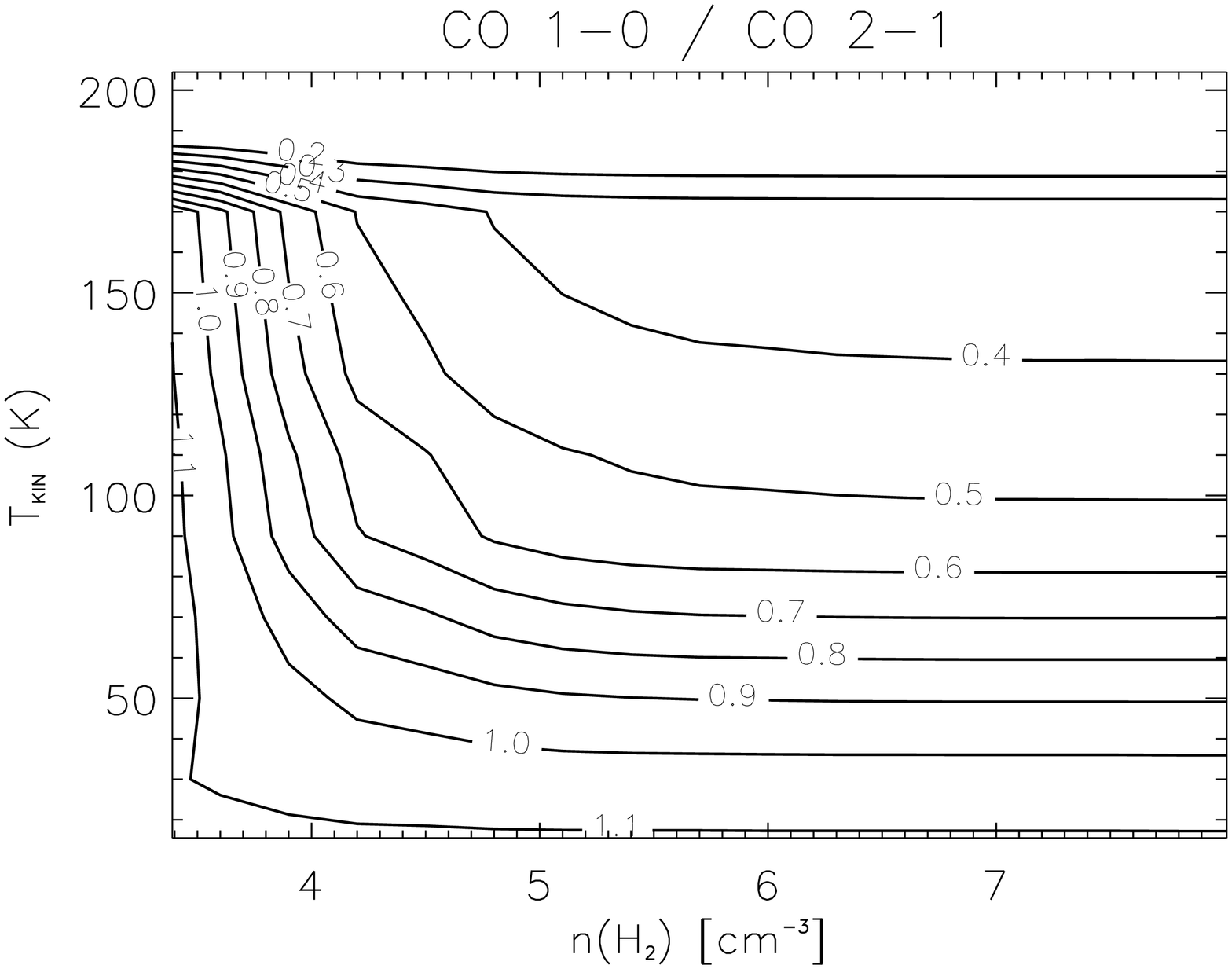} 
\caption{Line ratios of CO 1--0/CO 2--1 as a function of kinetic temperature and H$_{2}$ density for column density N$_{CO}$ = 10$^{18}$ cm$^{-2}$.}
\label{fig:radex_ratios}
\end{figure}

In the density range relevant for S~140 the CO 1--0/2--1 ratio is a good tracer for kinetic temperatures. This is demonstrated in Figure~\ref{fig:radex_ratios} showing the CO 1--0/2--1 line ratios for a CO column density of 10$^{18}$ cm$^{-2}$. For typical gas temperatures, we see that these ratios are insensitive to H$_{2}$ density for values $>$ 10$^{4.5}$~cm$^{-3}$, well above the critical density of both transitions. At this point we should also point out that at low kinetic temperatures this ratio is also not a good indicator of kinetic temperature at the accuracy needed to distinguish between 30~K and 60~K. In this temperature range CO 1--0/2--1 $\sim$ 1.0 $\pm$ 0.1, one would have to know reliably the main beam temperature to within 10 \% to trust the solution to the kinetic temperature to within $\pm$ 15~K. Given the uncertainties in antenna and receiver calibrations, beam sizes, sidelobes, etc. this level of accuracy is difficult to achieve. One main source of uncertainty is the coupling efficiency of the beam to the source structure. For sources larger than the main beam, the 30~m errorbeams (and sidelobes) contribute to the coupling. At 230~GHz, the main beam efficiency of the 30~m is high, 60 \%, and the three 30~m errorbeams contribute 26\% to the total power received by the beam pattern. In addition, assuming a reasonable source size of about 1 \arcmin, only the main beam and 1st errorbeam contribute. As the 1st errorbeam contains only 4\% of the total beam power, the correction factor for the antenna temperatures, Feff/Beff decreases  only  little, from 92/59=1.56 to 92/63=1.46. However, it is not the purpose of our paper to de--convolve the channel maps for the errorbeams or to provide a full accounting of the error budget contributing to T$_{mb}$.


\subsubsection{Assumptions}

For the CO 1--0, CO 2--1, $^{13}$CO and C$^{18}$O peak intensities ($\sim 3$$\times$~RMS) we performed a $\chi^2$ minimization to fit 
the gas kinetic temperatures and column densities
assuming that all lines arise from the same gas. For more accurate modeling the CO 2--1 map was convolved to a lower angular resolution in order to be consistent with the other lines ($\sim$21$\arcsec$). The $\chi^2$ function was computed as the quadratic sum
of the differences between the observed and the synthetic line intensities 
for a range of kinetic temperatures (10~K$<$T$_{\mathrm kin}$$<$200~K) and column densities (10$^{13}$cm$^{-2}$$<$N$_{CO}$$<$10$^{19}$cm$^{-2}$), values that are consistent with the expected ones for such region \citep[e.g.][]{vanderTak00,Spaans97,Poelman06,Huettemeister93}. 

An additional free parameter in RADEX is the temperature of the background radiation field that may pump
the line transitions. We adopted the value of 2.73~K for all our calculations since the cosmic background radiation (CMB) peaks at 
1.871~mm and it is the prominent component at millimeter wavelengths.
We used a fixed line width value of 3.5~km~s$^{-1}$ that approximates the value that we have measured throughout the cloud for the narrow component. 
Finally, in order to compute the intensities of the isotopic lines we assume fixed isotopic ratios of ${^{12}CO}/{^{13}CO} = 60$ and ${C^{16}O}/{C^{18}O} = 560$ \citep{Wannier80,Wilson94}.

\subsubsection{Gas density}
\label{sect:gas_density}

Previous studies have shown that the S~140 region contains gas with n$_{H_2}$ $\sim$ 10$^{5}$ cm$^{-3}$. \citet{Poelman06} reports clumpiness in the region giving a value of n$_{H_2}$ $\sim$ 10$^{4}$ cm$^{-3}$ for the interclump medium and~ $\sim$4$\times$10$^{5}$ for the clump gas.

\begin{figure}[ht]
\includegraphics[scale=0.43]{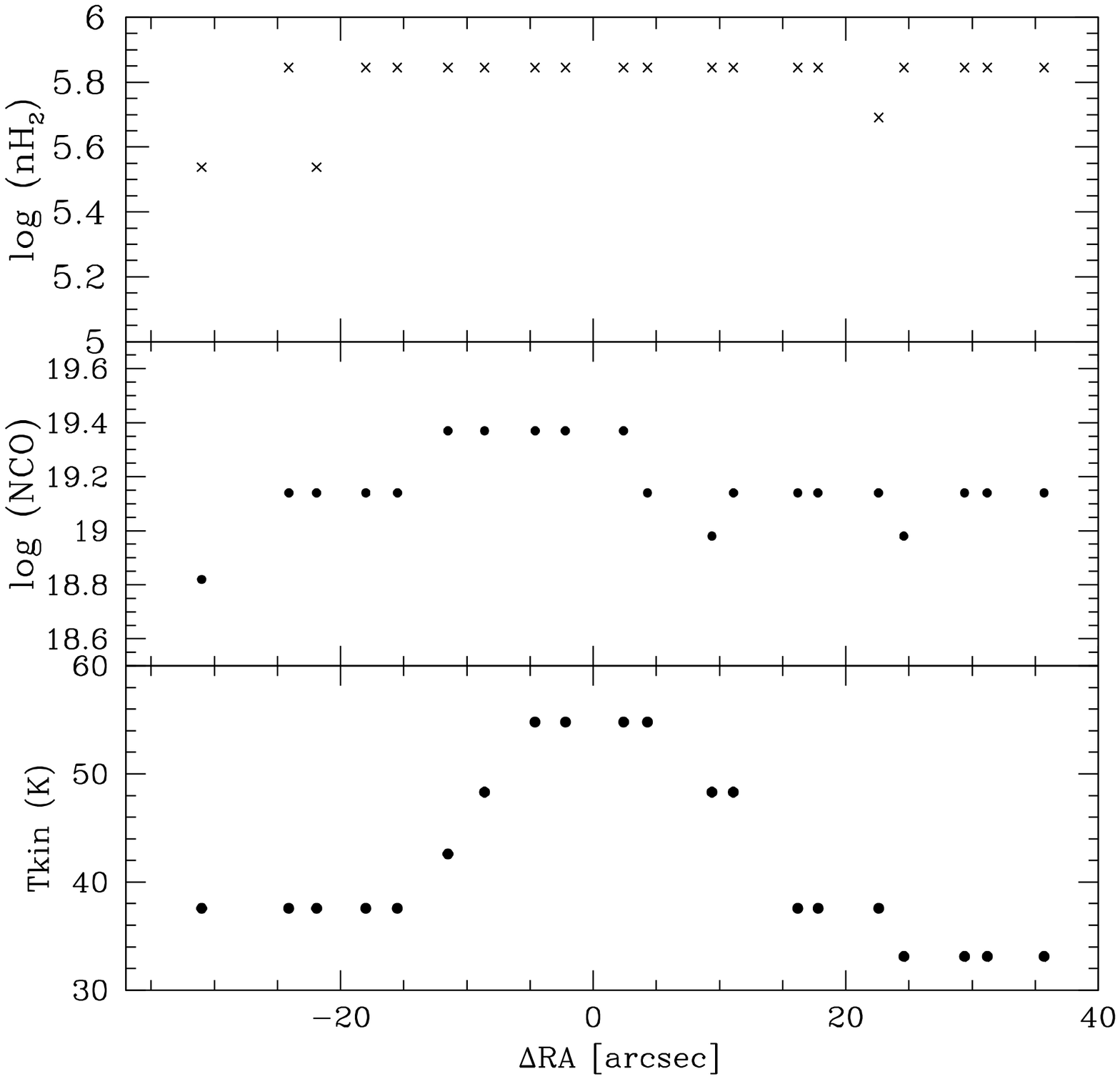}
\caption{Profiles of gas temperatures (bottom), column densities (middle) and n$_{H_2}$ densities (top) along the HIFI cut, as resulting from the 3--free parameter fitting process to the 30~m and HIFI data. The resulting n$_{H_2}$ densities are always $>$ 10$^{5}$ cm$^{-3}$ throughout the HIFI cut.}
\label{fig:3_plots_final.ps}
\end{figure}

Fig.~\ref{fig:radex_ratios} shows that it is impossible to derive the gas density from the low-$J$ lines of CO (2--1, 1--0) for densities of n$_{H_2}$ $\sim$ 10$^{5}$ cm$^{-3}$ or above. A reliable determination of the gas density was thus only possible for those parts of the map where the HIFI cuts provided additional high-$J$ line data. Here, we performed a three-parameter RADEX analysis, fitting the CO 1--0, CO 2--1, CO 9--8, $^{13}$CO 1--0, $^{13}$CO 10--9 and C$^{18}$O 1--0 lines convolved to the same resolution (21\as), to determine the n$_{H_2}$. The result is shown in Fig~\ref{fig:3_plots_final.ps}. This fit proves that, at least along the cuts, we find everywhere gas densities well above n$_{H_2}$ = 10$^{5}$ cm$^{-3}$, where the CO 1-0/CO 2-1 line ratio can be directly used as a temperature measure. In this way we can be sure to avoid the regimes where the ratio is sensitive to the H$_{2}$ density.

\begin{figure}[ht]
\begin{center}
\includegraphics[scale=0.43]{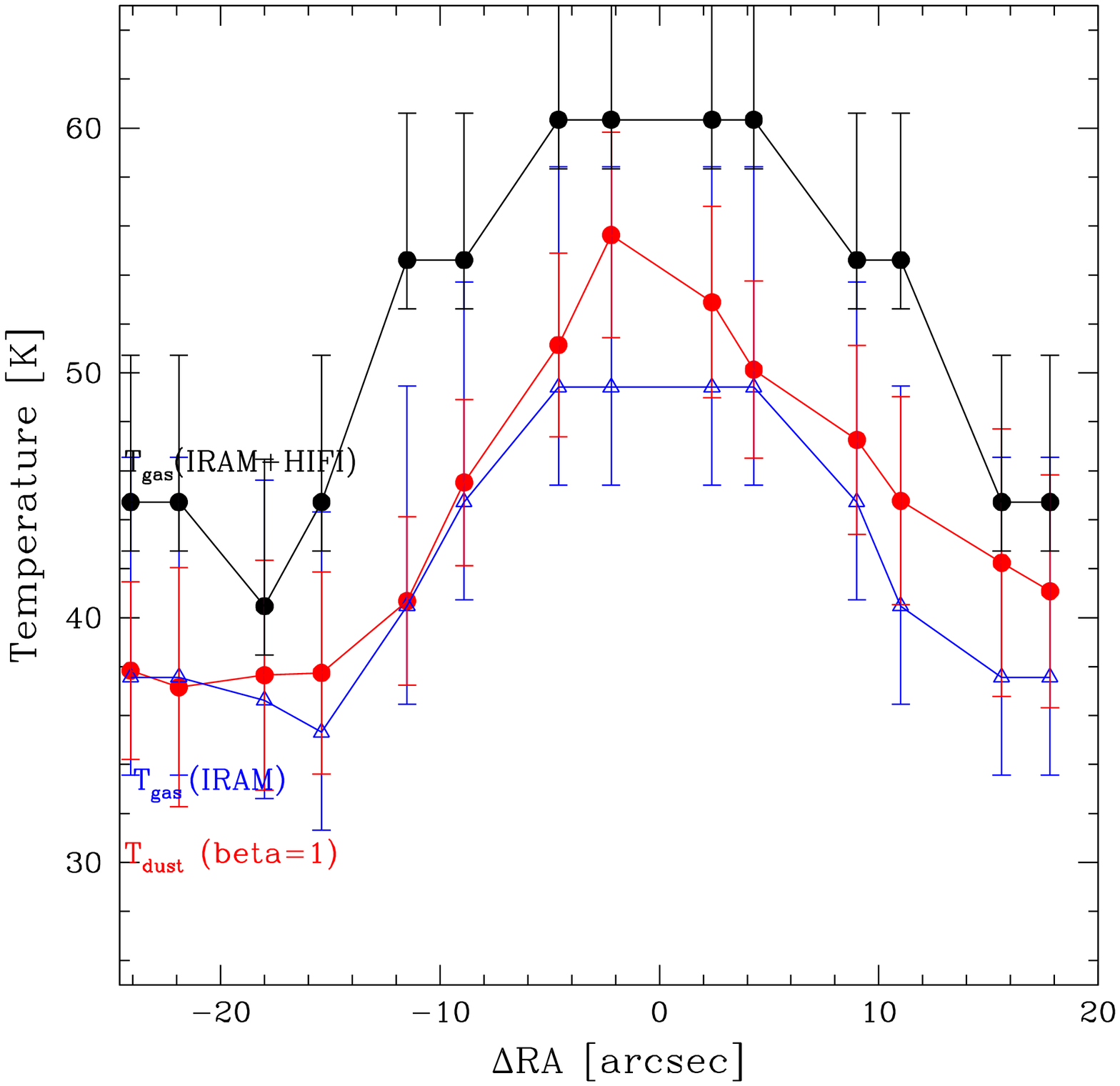}
\end{center}
\caption{Temperatures of dust (red line) and gas along the HIFI cut. The gas temperature as derived fitting low--$J$ lines (IRAM data, blue line) is systematically lower than the derived gas temperature fitting both low and high--$J$ lines (IRAM \& HIFI data, black line). T$_{\mathrm gas}$ (from both low and high--$J$ lines) is systematically higher than T$_{\mathrm dust}$, at least along this cut. The errors vary between $\sim$5--15~\% throughout the cut with the higher accuracy close to the central position.} 
\label{fig:Tkin_Tdust_cut.ps}
\end{figure}

To assess the reliability of the results, we perform two separate RADEX fits along the cuts where
the 9--8 and 10--9 transitions of CO and the isotopologues were observed by HIFI. In the first fitting approach
we fix n$_{H_2}$ to 10$^{5}$ cm$^{-3}$ and fit only the low--$J$ CO lines observed by IRAM (blue curve in Figure~\ref{fig:Tkin_Tdust_cut.ps}) and all the lines (red curve in Figure~\ref{fig:Tkin_Tdust_cut.ps}). The second fit uses all the lines and treats the gas density as a free parameter in the range between 7$\times$10$^{3}$cm$^{-3}$ and 7$\times$10$^{5}$cm$^{-3}$ as in Fig.~\ref{fig:3_plots_final.ps}. The best fit of the latter occurred for n$_{H_2}$ $>$ 10$^{5}$ cm$^{-3}$. Adding the higher--$J$ lines drives the fit to systematically higher kinetic temperatures by about $\sim$5--15~K (Figure~\ref{fig:Tkin_Tdust_cut.ps}). The opacities of the lines are found to be in the following ranges: CO 1--0: 16--45, CO 2--1: 52--154, CO 9--8: 3.6--65, $^{13}$CO 1--0: 0.3--0.8, $^{13}$CO 10--9: 0.017--0.38, C$^{18}$O 1--0: 0.025--0.09. As an attempt to test how much the optically thin $^{13}$CO 10--9 line influences the solution we re--ran our calculations applying a double weight for this line. The resulting kinetic temperatures changed by $\sim^{+2}_{-1}$~K that is inside the range of the reported errors.

We ran the same kind of analysis towards the two positions with most lines 
observed\footnote{For the low-$J$ CO data we extracted the points from the IRAM maps closest 
to HIFI pointing observations with no more than ~3$\arcsec$ offset.}
(see Table ~\ref{co_lines}).
Towards IRS~1 the full analysis provides a gas kinetic temperature of $\sim60^{+6}_{-2}$~K and a column density of $\sim3^{+1}_{-2}$$\times$10$^{19}$cm$^{-2}$ while the same procedure using only the IRAM data results in a kinetic temperature of $\sim49^{+9}_{-3}$~K and a column density of $\sim1.3^{+2}_{-2.2}$$\times$10$^{19}$cm$^{-2}$. Towards the IF position, the full dataset indicates
a kinetic temperature of $\sim40^{+5}_{-2}$~K and column density of $\sim1.0^{+2.5}_{-1.5}$$\times$10$^{18}$cm$^{-2}$, while the IRAM data only result in a kinetic temperature of $\sim35^{+9}_{-8}$~K and column density of $\sim4.4^{+3.6}_{-2.5}$$\times$10$^{17}$cm$^{-2}$. The gas temperatures obtained when using only the low-$J$ lines from IRAM underestimate T$_{\mathrm gas}$, providing a lower limit of gas temperatures \citep{Yildiz2013}. The major and minor axis of the resulted $\chi^2$ contours were used as the error bars of the two parameters.


For a more accurate determination of the H$_{2}$ density, tracers such as CS 2--1/3--2 and HCO$^{+}$ 1--0/3--2 are more reliable \citep{vanderTak07}, but were not observed. \citet{Snell84} derived a density of $5$$\times$10$^{5}$cm$^{-3}$ towards IRS~1--3 using four CS transitions which is in agreement with the value we derive. In addition \citet{Goldsmith99} performed a LTE population diagram analysis towards IRS~1 using the same CS dataset and derived a kinetic temperature between 30~K--50~K which is lower than our value probably due to the effect of their larger beam size.

\subsubsection{Kinetic temperature \& column density distribution}

\begin{figure}[ht]
\begin{center}$
\begin{array}{c}
\includegraphics[width=2.9in]{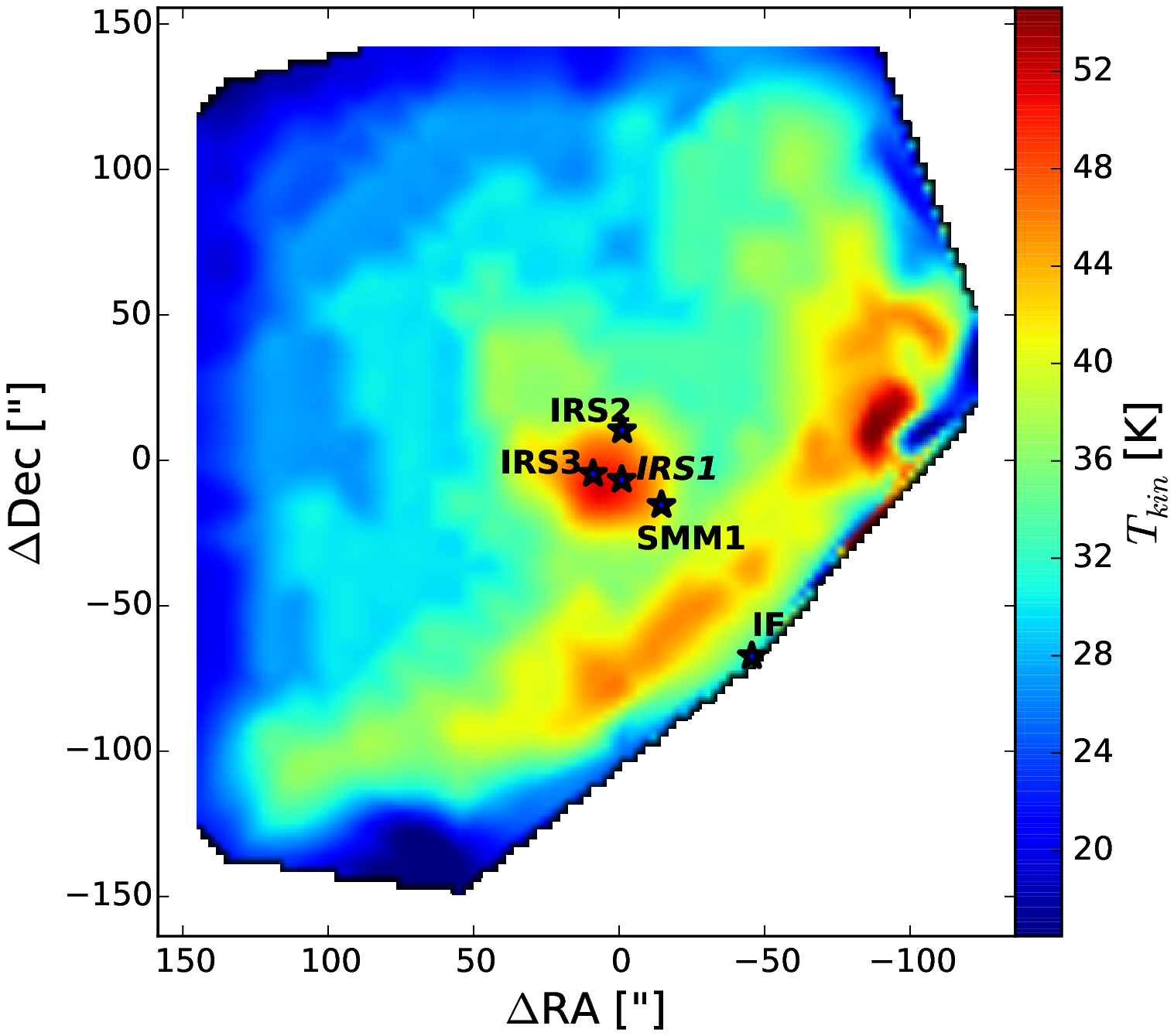}  \\ \includegraphics[width=2.9in]{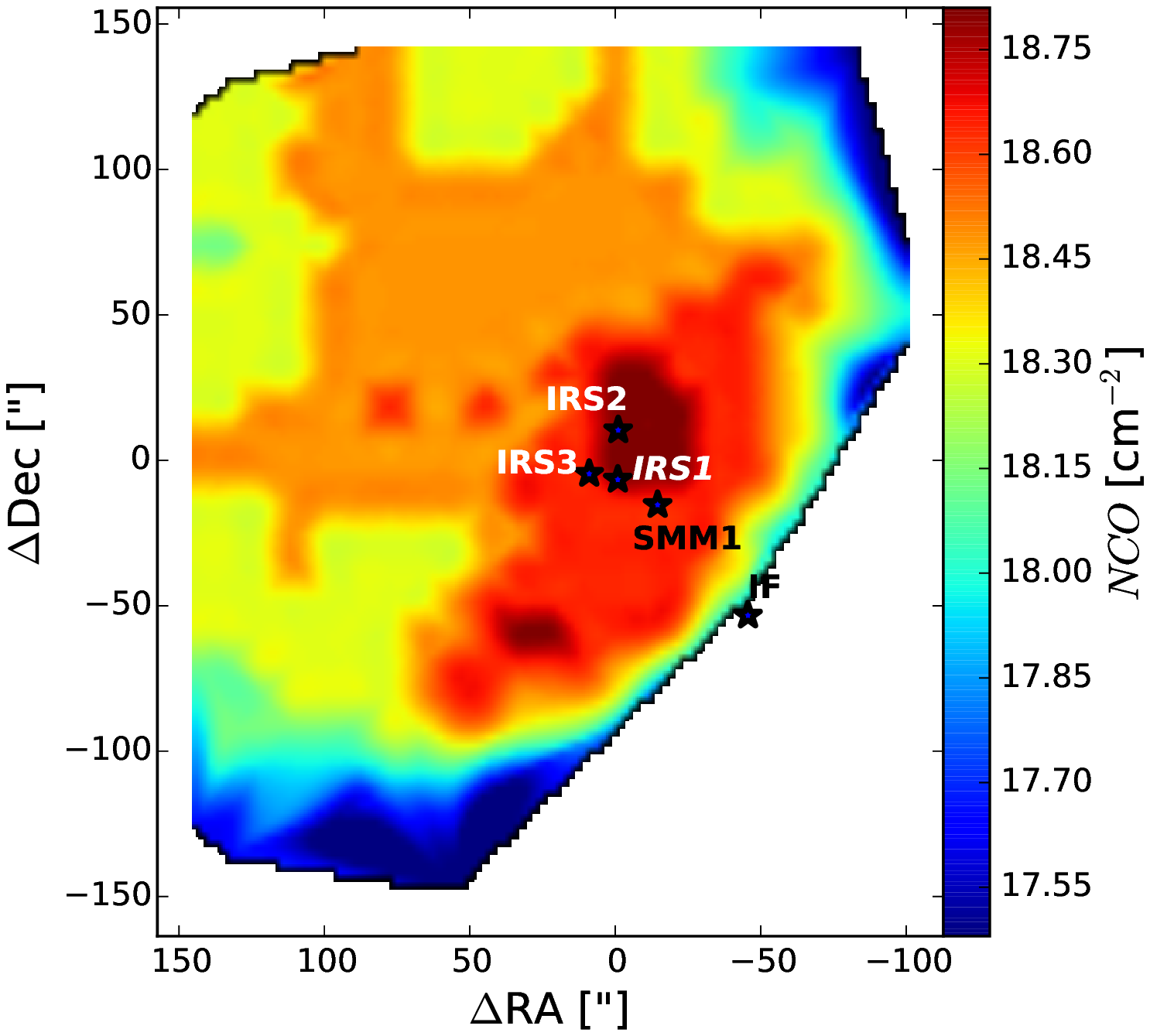}
\end{array}$
\end{center}
\caption{Kinetic temperature (top) and column density (bottom) maps of S~140 (all lines $>3$ RMS). The kinetic temperatures and column densities are higher towards the embedded infrared sources and lower throughout the rest of the cloud. The kinetic temperature rises again towards IF, but remains below the value towards the center.}
\label{fig:column_density_map_final.ps}
\end{figure}

The resulting maps of 
kinetic temperatures and CO column densities are shown in Figure~\ref{fig:column_density_map_final.ps}. 
We find a kinetic temperature of $\sim55$~K toward the center and $\sim45$~K 
toward the ionization front, while the rest of the cloud is characterized by lower temperatures (25--40~K) (Figure~\ref{fig:column_density_map_final.ps}). The column density toward the center was found to be the highest, with a value of $\sim6.5$$\times$10$^{18}$cm$^{-2}$, while toward the IF was found to be $\sim1.1$$\times$10$^{18}$cm$^{-2}$. The lowest column density that was determined throughout the cloud is $\sim1.4$$\times$10$^{16}$cm$^{-2}$. The gas temperature
map obviously reflects the different heating contributions, by the embedded 
cluster of the three high--mass YSOs (IRS~1 to 3) in the center and from the 
external B0V star HD211880 towards the IF.

\subsection{Dust Analysis}

\begin{figure}[ht]
\includegraphics[scale=0.46]{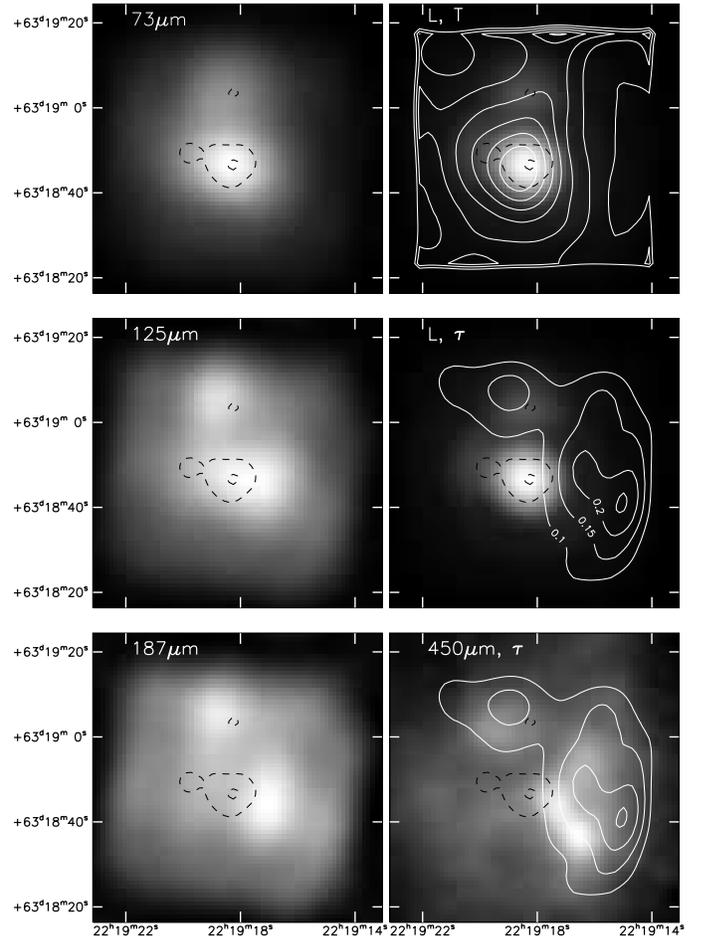}
\caption{Left column: PACS/Spec continuum images (5$\times$5 spatial array of
9\arcsec~pixels regridded to a 1\arcsec~grid) with
contours of 37$\mu$m (SOFIA) emission overlaid showing the positions of IRS~1, 2, and 3. Right column: top -- total luminosity image with contours of fitted dust temperature at 70, 65, 60, ...K; center -- luminosity image with
contours of fitted dust optical depth at levels of 0.1, 0.15, 0.2, and
0.25; bottom -- 450~$\mu$m SCUBA image with
same fitted dust optical depth contours overlaid.}
\label{fig:pacs}
\end{figure}

Prior to any radiative transfer modeling of the dust in this region (Sect.5) we here estimate
the rough temperatures and optical depths relevant to the dust distribution. For this we
used the PACS/Spec continuum data from 73 -- 187~$\mu$m which covers well the peak of the
SED over the entire region mapped with PACS (Figure~\ref{fig:pacs}).
To analyze our dust continuum observations, we used two subsets of the PACS and SOFIA images. To compute a luminosity
map, we used the 11 -- 187 $\mu$m images all as re-convolved to the 187~$\mu$m resolution (13\arcsec). At each 1\arcsec
pixel of the computed image we integrated the flux density from 11.1 -- 187~$\mu$m and included a
linear extrapolation to zero flux level beyond 187~$\mu$m, which typically resulting in only a few \% addition
to the total. We believe the absolute uncertainties in this map are of order 15\% based on
the absolute calibration uncertainties of all the input data. The relative uncertainties are
probably much less, $\pm$ 5\%, since there is very little luminosity shortward of 10~$\mu$m
and longward of 200~$\mu$m.

We computed color--temperature/optical depth values at each spatial point on the 1\arcsec~grid
by fitting a blackbody function modified by a $\lambda$ $^{-1}$ dust emissivity variation. The peak dust
optical depths at the shortest wavelength are likely several $\times 0.1$, but still less than unity,
so we did not include any effects from optical thickness in these estimates.
A comparison of the luminosity, color temperature, and optical depth maps and several
images is shown in Figure~\ref{fig:pacs}. These images show a number of important qualitative properties
of the dust emission. First, there is a clear and relatively smooth change in source morphology
from that at 73~ $\mu$m, which is similar to the 10--37~ $\mu$m images, to the morphology at 187~ $\mu$m
which is beginning to show many of the features of the SCUBA 450~ $\mu$m map. Secondly, the
dust optical depth we derived from the Herschel data is quite similar to the 450~ $\mu$m emission
map. Finally, the luminosity image shows clearly that IRS~1 is the principal luminosity
source in the region, followed in importance by IRS~2 and then IRS~3. There is no obvious
luminosity peak within the area of strongest 450~$\mu$m emission, so this is probably a peak in the column densities of dust and gas.

One uncertainty of the dust temperature may stem from the assumed spectral properties of the dust grains, in particular the assumed spectral index $\beta$. Recent observations of dust in the diffuse ISM with the Planck mission indicate a mean dust temperature of $\sim$20~K and dust emissivity index of $\sim$1.6 \citep{Jones2014}. This is clearly higher than the value of $\beta$$=$1 assumed here, that is more appropriate for dense clouds \citep{Ossenkopf94}. As our region is denser than the majority of those observed with Planck we can consider the value of $\beta$$=$1.6 as an extreme upper limit for our case. When we fit a blackbody function modified by a $\lambda$ $^{-1.6}$ to derive the dust temperatures we obtain temperatures that are lower by 5--20~K than the values from the adopted $\lambda$ $^{-1}$. We conclude that in our analysis we derive an upper limit of dust temperatures. 

\subsection{Comparison of Dust and Gas temperatures.}
\label{sect:Comparison_Dust_Gas}
\begin{figure}[ht]
\begin{center}$
\begin{array}{c}
\includegraphics[scale=0.35, angle = 270]{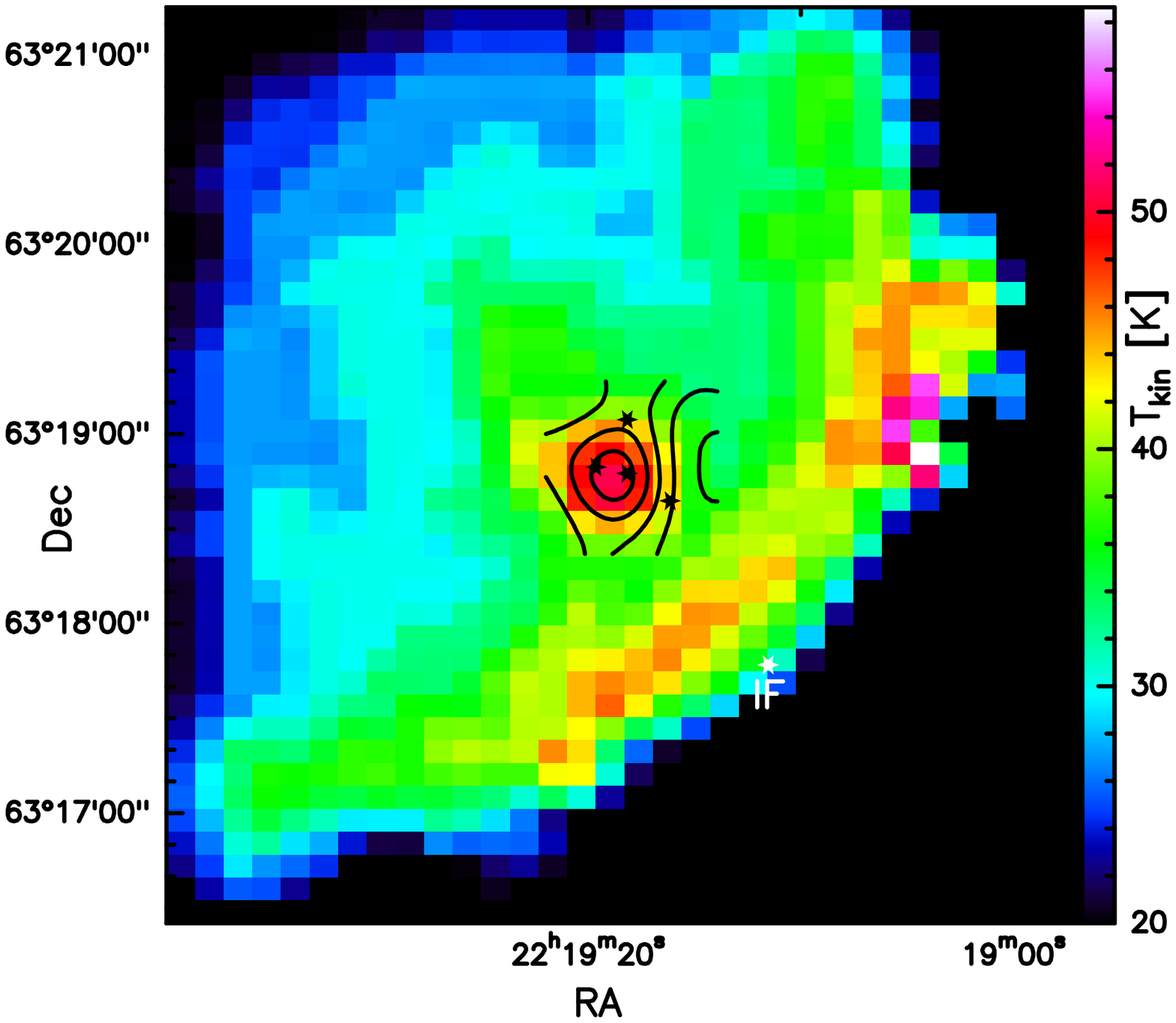} \\
\includegraphics[scale=0.35, angle = 270]{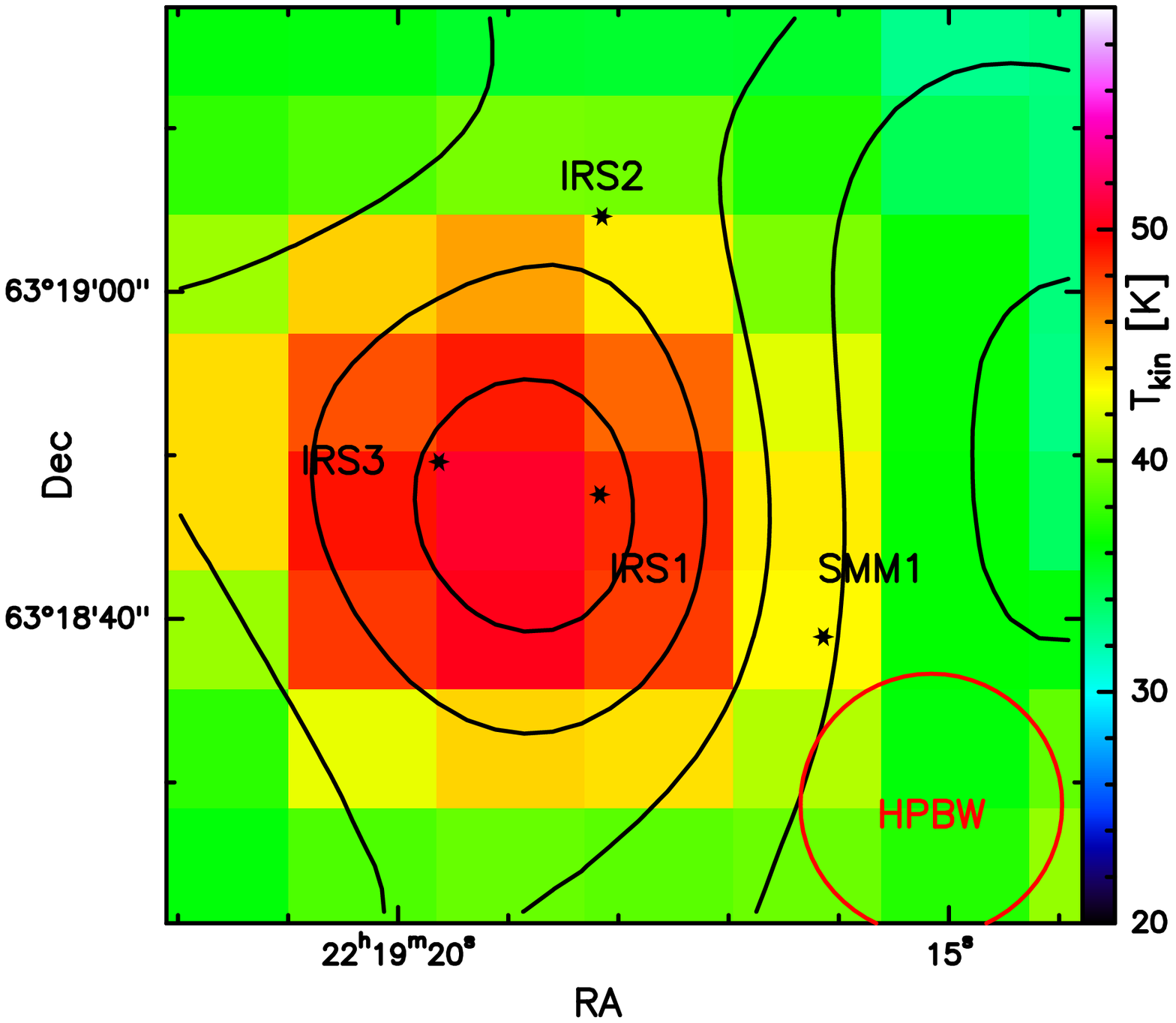}
\end{array}$
\end{center}
\caption{Gas kinetic temperature map towards IRAM field (low--$J$ analysis) with overplotted contours of dust temperatures (top) \& zoomed in PACS field (bottom). The 
levels of contours are 55~K (inner) till 35~K (outer) using a step of 5~K. The colour bar indicates the values of kinetic temperature. The 2 temperatures from this analysis seem to agree well but the gas kinetic temperatures are underestimated for $\sim$7--12~K (Figure~\ref{fig:Tkin_Tdust_cut.ps}).}
\label{fig:dust_gas_comparison.ps}
\end{figure}

For a direct comparison with the gas kinetic temperature map, the dust temperature map was convolved to the same angular resolution ($\sim$21$\arcsec$).
The comparison between gas and dust temperatures from the low--$J$ line analysis (Fig.~\ref{fig:Tkin_Tdust_cut.ps} \& Figure~\ref{fig:dust_gas_comparison.ps}), 
shows a very similar spatial structure around the infrared sources. The higher--$J$ lines from HIFI though, reveal that the low--$J$ analysis underestimates the gas temperatures (see Sect. \ref{sect:gas_density}). Assuming that this trend applies to the entire region and not only along the cuts (Figure~\ref{fig:Tkin_Tdust_cut.ps}) we conclude that the gas temperature is higher than the dust in the entire region.


The temperature difference of $\sim$5--15~K revealed from the complete analysis lies outside the uncertainty range of the 2--7~K and indicates a more efficient gas heating even at densities $\geq$ 10$^{4.5}$ cm$^{-3}$ where the two components are relatively well coupled.

\begin{figure}[ht]
\includegraphics[scale=0.37, angle=90]{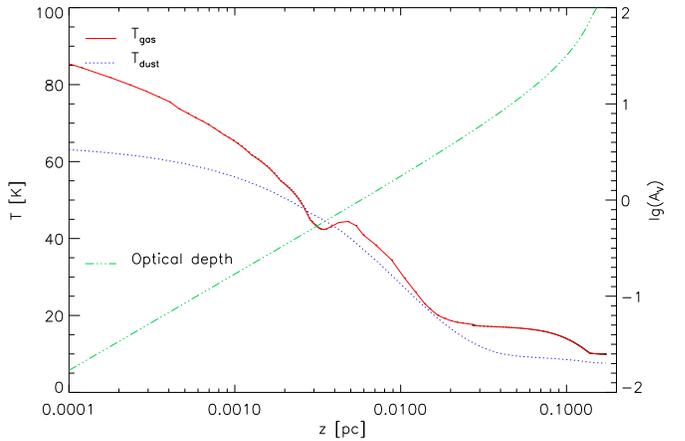}
\caption{Dust and gas temperatures in a spherical PDR clump with a surface density of $10^{5}$~cm$^{-3}$ illuminated by an UV radiation field of $10^{2}$ Draine fields. Gas temperatures generally exceed the dust temperatures \citep[see][for a more general discussion]{Roellig2013}. The green dash--dotted line shows the visual extinction $A_V$ as a function of the depth into the cloud.}
\label{fig:pdr}
\end{figure}

To illustrate the expected global temperature dependence, we show
in Figure \ref{fig:pdr} the outcome of a simplified PDR model for S~140 computed
with the KOSMA-$\tau$ PDR code \citep{Rollig2006,Roellig2013}. It shows
the temperature profile for a spherical PDR clump with a mass
of 100 $M_\odot$, mimicking a plane--parallel PDR, and a surface density 
of $10^5$~cm$^{-3}$ illuminated
by an UV radiation field of 100 Draine fields \citep{Draine1978}. For
reference, the green dash-dotted line shows the visual extinction
$A_V$ as a function of the depth into the cloud. The model shows that
throughout almost all of the whole cloud, the gas temperature is
higher than the dust temperature, by about 15~K for $A_V=0.1$ and by
only 2~K deep in the cloud. The small kink in the gas temperature
around $A_V=1$ stems from H$_2$ formation heating on PAHs and
the temperature structure of the PAHs. Overall the model confirms
that for a homogeneous medium, we expect the same temperatures for
gas and dust within our observational measurement errors for visual 
extinctions into the cloud of more than about 0.5, i.e. for depths of one 
arcsecond and more at the distance of S~140.
The extended hot gas therefore cannot be explained by a homogeneous
extinction of the UV radiation but requires some clumpy structure
allowing for a deeper UV penetration.

\section{Effect of density gradients}

Assuming uniformly distributed densities and temperatures for each spatial position was a good approach for the extended gas and dust in S~140, but a detailed modeling is required for the more complicated dust and gas structures that are expected to be close to the infrared sources. The radiation we receive is a result of various processes including the interaction with the surrounding dust. For a more precise physical approach, the radiative transfer calculations should include the original radiation, characteristics of the dust grains and the dust density distribution. Density gradients have been revealed for S~140 in previous studies including \citet{Harvey78}, \citet{Guertler91}, \citet{vanderTak00}, \citet{Maud13}. We first perform an advanced dust modeling taking into account dust density gradients. Then we apply the best--fit derived dust model in order to predict the CO intensities using an advanced radiative transfer code \citep[i.e. RATRAN;][]{Hogerheijde2000}, as an attempt to test the accuracy of our dust model comparing with the gas observations and predictions. 

\subsection{Dust modeling -- Approach}

It is clear from all the infrared and sub--mm images of the S~140 cluster that there are a number of
luminosity sources in the central arcminute {\it and} that the dust distribution is unlikely to be
spherically symmetric. This implies that any simple model of the dust heating will be limited
in its applicability. We have, however, identified some goals
to try to address in this study with simple radiative transfer models. First, since all the observational data
suggest that IRS~1 is the dominant luminosity source, we believe that we can determine some
rough properties of the dust distribution close to it by ignoring heating from IRS~2 and 3 and any
of the other much lower luminosity objects that are part of the embedded stellar cluster around
IRS~1. To this end, for the model comparison for IRS~1 we concentrate on two observables: (1) the SED of IRS~1, and
(2) the spatial flux distribution from the center to the south of IRS~1 where there are no strong
obvious nearby heating sources seen in the infrared that are likely to contribute significantly 
to the dust heating. A second goal is to try to understand whether the
peaks in the 450~$\mu$m
image can be due to some unusual dust distribution to the west and southwest of IRS~1 with no additional internal
heat source or if some internal heating is required to explain them. For example, the recent observations and modeling by
\citet{Maud13} suggest an internal source within the strongest 450 $\mu$m peak to the southwest of
IRS~1, i.e. SMM~1 ($\Delta$RA: -14.22\arcsec, $\Delta$Dec: -7.90\arcsec, relative to IRS~1).

The approach we took to addressing these goals involved several steps which we discuss in detail in the
following sections. First, we tried to find what properties of models were required to provide a
fit to the observed spectral energy distribution (SED) from the central pixel centered on IRS~1 and 
to the source profile of IRS~1 to the south, assuming a
single, spherically symmetric dust distribution. This model was then used as a starting point for
subsequent stages.
The second step was to create a model made from hemispheres of two different
spherically symmetric dust distributions. Though this model is simplified, it seems likely that far from the
boundary region, such a combination can give insight into the degree to which such a dust distribution
might reproduce the observed source structure to the west and southwest of IRS~1 by assuming a higher
column density in this region than to the south and east of IRS~1. 
For this comparison we used the source profile extending southwest along a line from IRS~1 to and beyond
the position of SMM~1 as given by \citet{Maud13}.
As an alternative to the two-hemisphere model, we also attempted to fit the profiles along this line with two separate sources by
using the superposition of two spherically symmetric models with different central source luminosities
and very different dust distributions separated on the sky by $\sim$ 16\arcsec~to attempt to reproduce
the observed 1-D flux profile between IRS~1 and the position of SMM~1. 
southwest of it.

\begin{figure}[ht]
\includegraphics[scale=3.5]{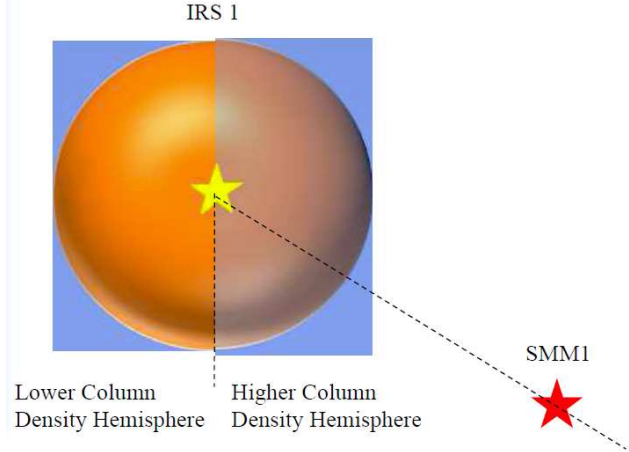}
\caption{A schematic diagram in the place of the sky of the model for IRS~1 and the submm peak ($\Delta$RA: -14.22\arcsec, $\Delta$Dec: -7.90\arcsec) to the southwest. The distance between the 2 positions is $\sim$~0.06~pc.}
\label{fig:schematic}
\end{figure}

In testing these last two model variations, the two-hemisphere construction and the two-source construction, it became clear that the best results would likely result from a combination of both features. Such a model
would also be most realistic in light of the much higher dust column density to the west and southwest as
indicated both by the SCUBA maps {\it and} our {\it Herschel} maps, together with the presence of the compact
sub-millimeter/millimeter source found by \citet{Maud13}. After constructing a few such models,
we realized that with so many free parameters, it would be difficult to find the range of well-fitting
parameters without testing the fits over a large, multi-dimensional grid. This process represented the final stage of
modeling for IRS~1, and we now describe the process and results. Figure \ref{fig:schematic} shows a schematic diagram
of the model for IRS~1 and the submm peak to the southwest.

For all the models we used the DUSTY code \citep{Ivezic97} and converted the dimensionless output
values to those appropriate for the assumed distance of S~140, 746~pc \citep{Hirota2008} and the 
assumed luminosity (discussed below) for comparison with the observations.
We tried models with two different dust compositions; the first was a mixture of
90\% DL silicates \citep{Draine84} and 10\% amorphous carbon, both of whose optical constants are
distributed with DUSTY. The second was OH5 \citep{Ossenkopf94} dust which a number of authors \citep[e.g.][]{vanderTak1999} have suggested
may be more appropriate in dense star-forming regions where significant grain growth may have taken place. 
We found similar quality fits with comparable dust density distributions for both grain compositions. 
The 1 $\mu$m optical depths were, however, of order twice as high with the DL/carbon dust than with
the OH5 grain properties
due to the overall difference in the slope of the dust optical depth between 1 and 100 $\mu$m.
The output of the ``observed'' profile of each model at each wavelength was first interpolated onto a 
3050$\times$850 grid with spacing equivalent to 0.05\arcsec. We then convolved this image with
the appropriate observing beam to compare with the observed images (not those convolved to the
187 $\mu$m resolution). Finally, we ``observed'' this image with a 9\arcsec~square beam that we
moved across the image in 1\arcsec~steps to compare with the observed images whose fluxes we also
added into such a moving 9\arcsec~square beam. For the shorter wavelength images we also applied the
same 9\arcsec~square pixel flux summation for comparison between the models and observations.

\subsection{Model Grid}\label{grid}

The model that we used is described by 9 parameters, four for the eastern half of IRS~1, four for the western half, and
one, luminosity, for SMM~1.
Table \ref{gridtbl} lists these 9 parameters as well as the range of values tested in a grid allowing all possible
combinations of these values (150,000 models). A preliminary run of $\sim$ 800,000 models showed that the results were quite insensitive to
the value of parameter 4, the radius where the slope of the density gradient changed in the high-column hemisphere, so it was
fixed at a value of 300 $\times$ the inner radius. Similarly,
the value of the density gradient in this inner region affected the final results only slightly, so this was fixed at a value 0.4 ($\rho \propto r^{-0.4}$).
Therefore, for the final results there are seven free parameters.

\begin{table}[ht]
\caption{Model Parameters And Range for Grid Fitting With OH5 Dust.}

\centering
\small\addtolength{\tabcolsep}{-3.0pt}
\begin{tabular}{c c }
\hline\hline
Parameter & Value or Range \\
\hline\hline
{\bf Hot Source (IRS1) 1.0$\times 10^4$\lsun} \cr
Density Gradient, r$^{-p}$ & 0.0 -- 0.9\tablenotemark{a} \cr
Grain Size Distribution Slope & 3.5 \cr
Grain Size Min/Max ($\mu$m) & 0.1/1.0 \cr
Dust Temp at Inner Radius & 1400K at 2.0$\times 10^{14}$ cm \cr
R$_{outer}$/R$_{inner}$ & 750 -- 3000\tablenotemark{a} \cr
A$_v$ & 20 -- 90 mag\tablenotemark{a} \cr
\cr

\bigskip
{\bf Cold Source (SMM1) } \cr
Luminosity &  0 -- 300\lsun\tablenotemark{a} \cr
Photosphere Temperature & 2500 K \cr
Dust Temp at Inner Radius & 1200K \cr
Density Gradient, r$^{-p}$ & 1.0 \cr
R$_{outer}$/R$_{inner}$ & 3000 \cr
Grain Size Distribution Slope & 3.5 \cr
Grain Size Min/Max ($\mu$m) & 0.1/1.0 \cr
A$_v$ & 1000 mag \cr
\cr

\bigskip
{\bf Southwestern Profile} \cr
Inner Density Gradient, r$^{-p}$ & 0.4  From R$_{in}$ to 300 R$_{in}$  \cr
Outer Density Gradient, r$^{-p}$ & 1.5 -- -2.0 From 300R$_{in}$ to
R$_{out}$\tablenotemark{a}  \cr
R$_{outer}$/R$_{inner}$ & 750 -- 3000\tablenotemark{a} \cr
Grain Size Distribution Slope & 3.5 \cr
Grain Size Min/Max ($\mu$m) & 0.1/1.0 \cr
Dust Temp at Inner Radius & 1400K \cr
A$_v$ & 15 -- 70 mag\tablenotemark{a} \cr
\hline
\tablenotemark{a} Free parameter during modeling.

\end{tabular}
\label{gridtbl}
\end{table}

In order to select the most likely models and to estimate the range of
values that produce a reasonable fit, we characterized the model fits by the $\chi^2$ that we computed as
follows. We parameterized the observations
and accompanying model fits into 30 values of the SED and the source profile at various wavelengths. 
Since the signal-to-noise ratio of almost all the observations is quite
high, the systematic uncertainties of the observations (e.g. pointing) dominate the true uncertainties. We, therefore, assigned somewhat 
arbitrary values to the assumed errors for these 30 values to drive the fitting to
something ``reasonably'' close to the observations. 
Table \ref{gridvaltbl} lists these properties of the observed
data and uncertainties.

\begin{table}[ht]
\caption{Parameters and Assigned Relative Uncertainties Used to Compute $\chi^2$ of Fit To IRS~1.}

\centering
\small\addtolength{\tabcolsep}{-5.0pt}
\begin{tabular}{l c }
\hline\hline
Parameter(s) & Relative Uncertainty \\
\hline\hline
Peak F$_\nu$ at 37, 450$\mu$m & 0.1 \cr
Peak F$_\nu$ at 73, 125, 187$\mu$m & 0.05 \cr
SMM~1 F$_\nu$ at 450$\mu$m & 0.1 \cr
Relative F$_\nu$ at 37, 73, 125, 187, 450$\mu$m~\tablenotemark{a} & 0.1 \cr
Relative Luminosity~\tablenotemark{a}& 0.1 \cr
\hline
\tablenotemark{a} The offsets were at -18\as, -9\as, 9\as and 18\as as appear \\ along the dashed line in Fig~\ref{fig:schematic}.
\end{tabular}
\label{gridvaltbl}
\end{table}

As shown in the table, the parameters used to calculate the $\chi^2$ of the fit included: the
SED between 37 and 450~$\mu$m, the flux at the position of SMM~1, the relative flux at 37--450~$\mu$m at offsets of 9\arcsec~and 18\arcsec
to either side of the position of IRS~1 (i.e. one pixel and two pixels), and finally, the observed luminosity at those four offset
positions (not relative to the central pixel). We intentionally did not include the portion of the SED shortward of 37~$\mu$m, since the dust close to
the central source of IRS~1 is likely to be distributed in a disk-like configuration which would lead to more flux
escaping at shorter wavelengths than consistent with the assumption of spherical symmetry. Likewise, we did not attempt
to fit the total observed luminosity at the central pixel since much of the luminosity is emitted shortward of 37~$\mu$m.
To estimate the range of model parameters that provided a ``good'' fit to the data, we
assigned probabilities to each model equal to ${\rm exp}(-\chi^2/2)$. 




The best--defined parameter values from the probability plots, 
are the mild density gradient away
from IRS~1 of $\rho \propto r^{-0.4}$, and the outer radius of 1500 times the inner radius. The optical depth is relatively well defined by
A$_v \sim$ 40, though there is a better--defined joint probability that includes the optical depth and density gradient. The peak
probability for the luminosity of SMM~1 is 100\lsun, but there is a wide range of values with reasonable probability from $\sim$ 50\lsun\ to 300\lsun.
The parameters of the dense material to the southwest of IRS~1 are not well defined for the most part. Interestingly the gradient of the material
in the outer area (300 -- 1500 $\times$ the inner radius) seems to have a most probable range similar to that found for the rest of the
cloud around IRS~1, i.e. $r^{-p}$ with $p \sim$ 0 -- 0.5. 
There is a significant joint probability distribution between the outer gradient to the 
southwest and the
luminosity of SMM~1. This is to be expected, since the effects of a steeper gradient can be largely compensated by increasing the luminosity
of SMM~1. Although the most probable models have A$_v$ = 40, the probability distribution suggests that the most probable range is 40 -- 60,
i.e. somewhat higher than the gradient to the south.

\begin{figure}[ht]
\includegraphics[width=3.in]{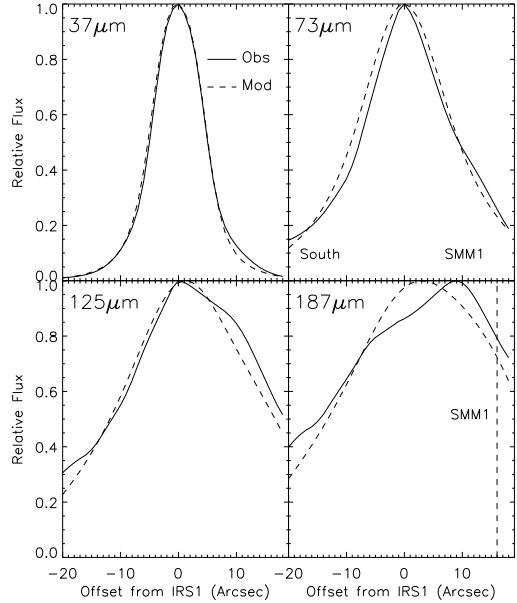}
\caption{Observed and model profiles at four PACS/Spec wavelengths
normalized to the peak along a line
from the south into the center of IRS~1 and then out along a line to the
southwest extending to the
SMM~1 source.}
\label{bestchip}
\end{figure}
\begin{figure}[ht]
\includegraphics[width=3in]{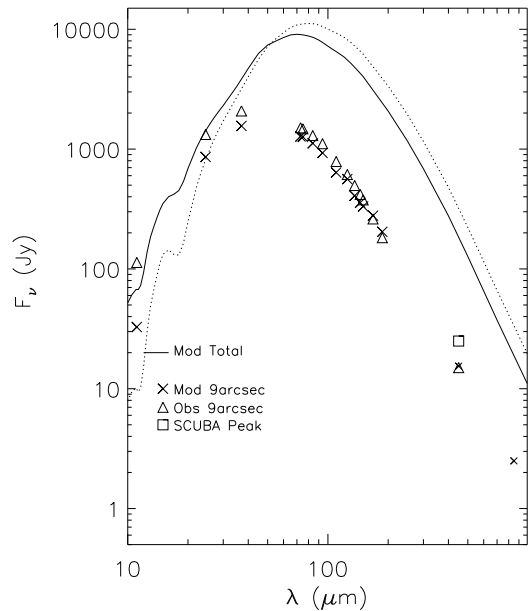}
\caption{SED for a typical model fit with low $\chi^2$. The solid line
shows the total flux coming from the
warm hemisphere, and the dotted line shows the SED for the fraction of
the flux coming from the cold hemisphere.
The plotted symbols show the observed and model fluxes for the fraction
within the central 9\arcsec~PACS pixel.}
\label{bestchis}
\end{figure}
\begin{figure}[ht]
\includegraphics[width=3.2in]{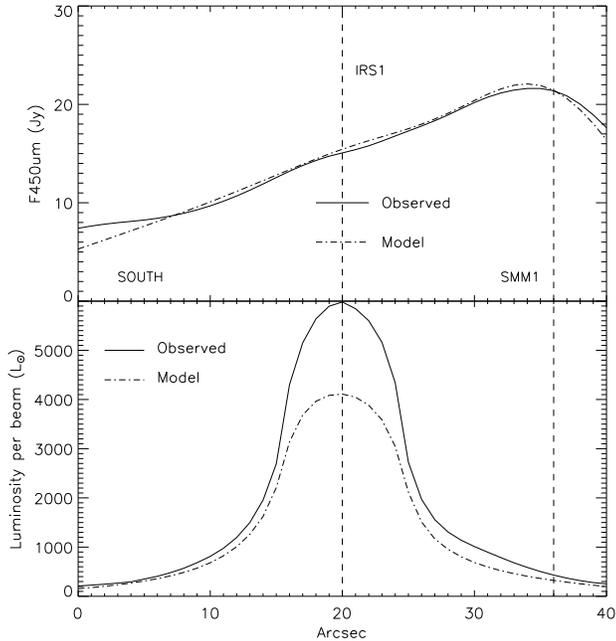} 
\caption{Profiles along the same line used in Fig.~15 for the 450~$\mu$m flux (top panel) and the total luminosity
(bottom panel). Note the significant mismatch in the central region
which is due to not trying to fit
the 5-30~$\mu$m luminosity since some of it likely arises from an
axisymmetric disk rather than the
spherical model used here.}
\label{bestchil}
\end{figure}

Examples of the model fit profiles from one of the two lowest $\chi^2$ (2.7) fits as well as the SED are shown in Figures \ref{bestchip}--\ref{bestchil}.
The model profiles shown in Figure \ref{bestchip} display a less than perfect fit at the longest wavelengths in our PACS observations, 140 -- 190~$\mu$m.
This effect has been driven by two parts of the fitting process, the fact that we did not include the relative flux at the central (IRS~1) position
in the $\chi^2$ and the fact that it proved quite difficult to find any models that dropped as slowly as the observations at the ends of
the profiles 18\arcsec~away from the center. The properties of the models that do the best job fitting the profiles 18\arcsec~away from IRS~1 also
drive the relative 187~$\mu$m profile to be too high at IRS~1 compared to its value at SMM~1. In some simple tests we have found that 
if we added an artificial floor to the model fluxes beyond
125~$\mu$m, we could fit the longer wavelength profiles significantly better. 


\subsection{IRS~2 and 3}\label{irs23}

\begin{figure}[ht]
\includegraphics[width=3.in]{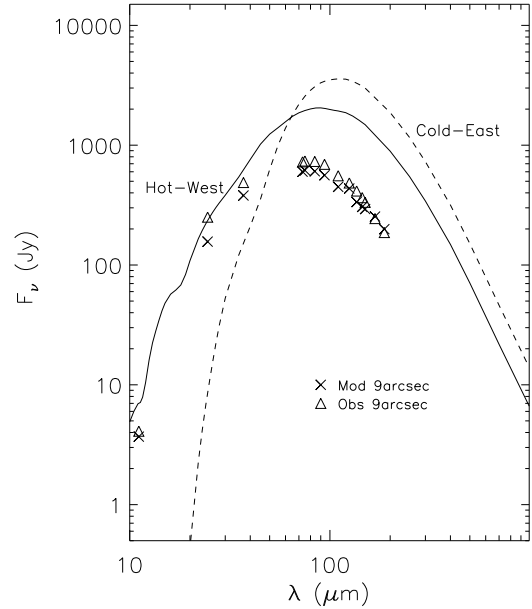}
\caption{Model and observed SED's for IRS~2. The solid line shows the
model fluxes for the hotter source to
the west in the model, while the dashed line shows the fluxes for the
colder source to the east. The symbols show the model and observed fluxes in the central 9\arcsec~pixel as in Fig.~16.}
\label{irs2s}
\end{figure}
\begin{figure}[ht]
\includegraphics[width=3.in]{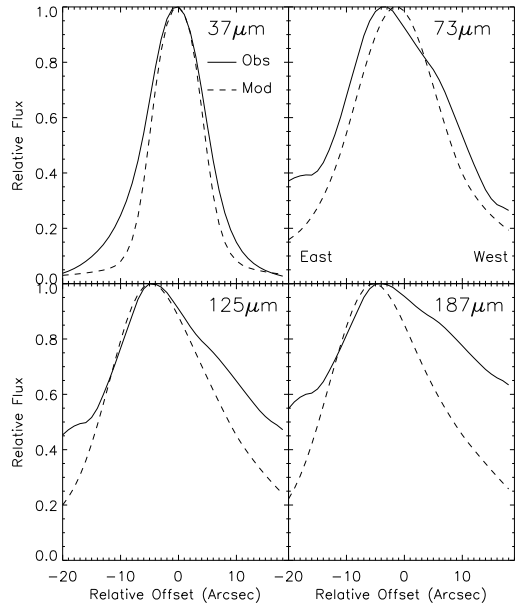}
\caption{Model and observed profiles for IRS~2 as in Fig.~15, but along a straight east--west line.}
\label{irs2p}
\end{figure}
\begin{figure}[ht]
\includegraphics[width=3.in]{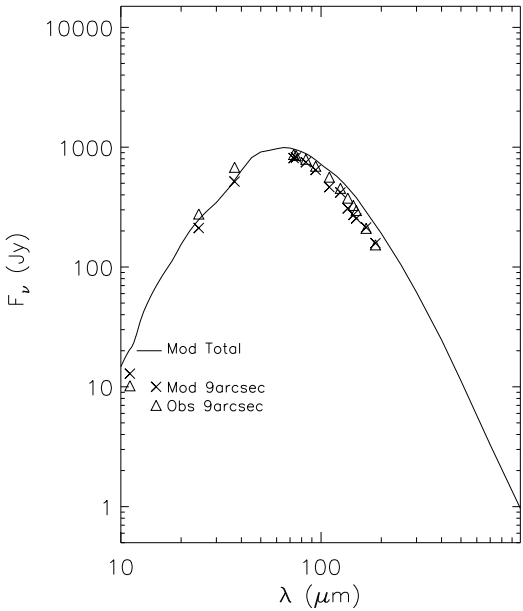}
\caption{Model and observed SED´s for IRS~3.}
\label{irs3s}
\end{figure}
\begin{figure}[ht]
\includegraphics[width=3.in]{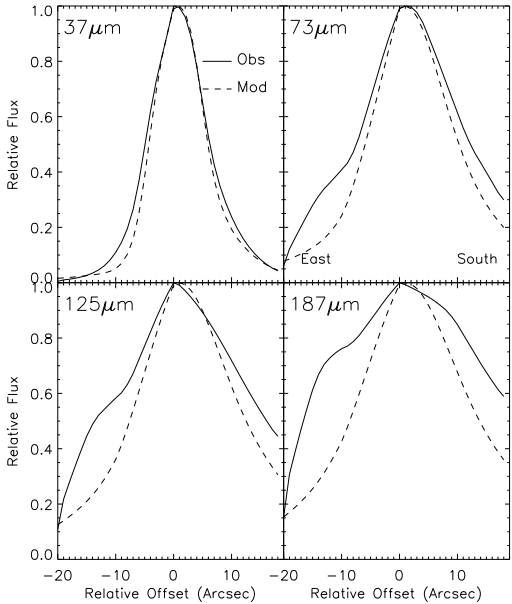}
\caption{Model and observed profiles for IRS~3 along a line extending into the source center from the east, and then outward to the south.}
\label{irs3p}
\end{figure}

The most important factor affecting our attempts to model IRS~2 and 3 is that the flux from our well-fitting models
for IRS~1 is a significant fraction of the total flux at the positions of IRS~2 and 3. This means that any models we
make for IRS~2 and 3 will have an additional large uncertainty due to the uncertainty in the true contribution from
IRS~1. For example, the model flux from IRS~1 at the position of IRS~3 is more than 50\% of the total at wavelengths
beyond 120~$\mu$m. We have therefore computed models for roughly a dozen possible configurations for each source
to develop some feeling for the range of likely parameters, but the uncertainties in our estimates are quite large.
For IRS~3 we used a profile that begins on the east side of the source, and after passing through the center, goes
straight to the south in order to minimize the confusion from IRS~1.
An additional complication is that IRS~2 is clearly elongated in the east-west direction in the mid-IR, and there is
a clear position shift in the same direction between the mid-IR and our PACS data. This suggests that there are
at least two relatively luminous objects heating the dust at the position of IRS~2 with different dust optical depths,
higher to the east and lower to the west. The position shift between 12 and 73 ~$\mu$m is only a few arcsec, which
is a fraction of our PACS native pixel size. Therefore, for IRS~2 we used a source profile from east to west and a
model like that in the ``two-hemisphere'' modeling for IRS~1, with a hot, lower-optical-depth dust distribution on
the western side and a cold, higher-optical-depth distribution offset 4\arcsec~to the east. Such a model will not
accurately reproduce the source elongation, but should roughly fit the overall SED and some part of the profiles.
Figures \ref{irs2s}--\ref{irs3p} show the model SED and profile fits for some representative models that illustrate
both the features that can be fit reasonably well as those that are difficult to fit. 

For both of these sources we
{\it did} attempt to fit the mid-infrared fluxes in the SED, unlike for IRS~1.
The very extended tail of emission on the west side of IRS~2 at 125 and 187~$\mu$m coincides well with the increase
in optical depth seen in our data and the SCUBA maps in that area. We have not tried to fit this with these simple
models. The bump seen on the eastern side of the long wavelength profiles for IRS~3 is coincident with the outermost
pixel of our mapping, and we have likewise not attempted to fit it.

The most well-defined results seem to be that: 1) the dust distribution around both sources is likely to be quite
flat, and 2) the luminosities of the sources are likely in the range of 1000 -- 2000\lsun, as shown in Table \ref{irs23fits}. The first could
be consistent with very little dust around them that is associated with them, but rather mostly with
the extended distribution around IRS~1.

\begin{table}[ht]
\caption{Model Parameters For Symmetric Fit With OH5 Dust to IRS~3 and ``Two-Hemisphere'' Fit To IRS~2.}

\centering
\small\addtolength{\tabcolsep}{-3.0pt}
\begin{tabular}{l c c }
\hline\hline
Parameter & IRS~2 & IRS~3 \\
\hline\hline
\bigskip
{\bf Single Gradient} & Eastern Side & Symmetric \cr
Source Luminosity & 2000 \lsun\ & 1300 \lsun\ \cr
Dust Temp at Inner Radius & 1400K & 1400K \cr
Density Gradient, r$^{-p}$ & 0.0 & 0.0 \cr
Grain Size Distribution Slope & 3.5 & 3.5\cr
Grain Size Min/Max (~$\mu$m) & 0.1/1.0 & 0.1/1.0 \cr
R$_{outer}$/R$_{inner}$ & 3000 & 1500 \cr
A$_v$ & 30 mag & 20 mag \cr
\cr

\bigskip
{\bf Double Gradient} & Western Side & \cr
Inner Density Gradient, r$^{-p}$ & 0.3 From R$_{in}$ to 300 R$_{in}$ & \cr
Outer Density Gradient, r$^{-p}$ &  0.3 From 300R$_{in}$ to R$_{out}$ & \cr
R$_{outer}$/R$_{inner}$ & 1500 &\cr
A$_v$ & 110 mag & \cr
\hline
\end{tabular}
\label{irs23fits}
\end{table}

\subsection{Gas modeling}

\begin{figure}[ht]
\includegraphics[scale=0.5]{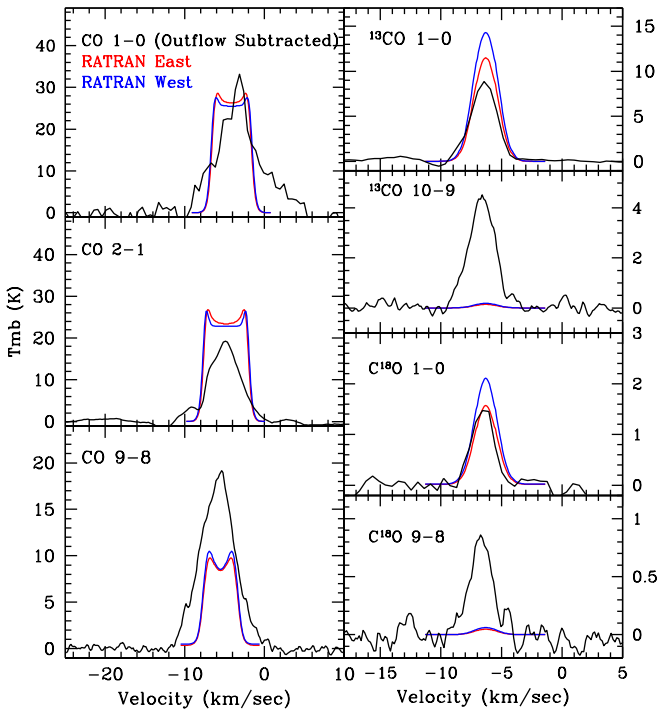}
\caption{Outflow subtracted spectra of CO 1--0, CO 2--1, CO 9--8, $^{13}$CO 1--0, $^{13}$CO 10--9, C$^{18}$O 1--0 and C$^{18}$O 9--8 (black) as observed towards IRS~1, overplotted with the synthetic emission as calculated with RATRAN having applied the lower density east hemisphere from DUSTY models (red) and the high density west hemisphere (blue). We observe no significant differences between the two cases with the exception of the low--J transitions of the isotopologues. While the model fits well the observations for the lower transition lines, it fails to reproduce the observed higher transition lines.}
\label{fig:ratran}
\end{figure}

In order to test our best fit DUSTY model we ran the Monte Carlo radiative transfer code RATRAN \citep{Hogerheijde2000} which treats the physical structure of the sources including temperature and density gradients. RATRAN estimates the local radiation field at all line frequencies taking into account the radiation field from every other position in the cloud. We ran RATRAN applying the best fit DUSTY model and assuming the same temperature for dust and gas towards IRS~1 in order to model the CO 1--0, CO 2--1, CO 9--8 and isotopologue lines and compare them to the outflow--subtracted component from the observations (Figure~\ref{fig:ratran}). The outflow emission was subtracted since the DUSTY models focus on the bulk of the quiescent dense gas where the outflow cavities are negligible. For the outflow subtraction we applied a 2 component Gaussian fit to the observed lines towards IRS~1 followed by the subtraction of the broad component.

For our calculations we defined a grid of 18 spherical shells for the east (lower density) hemisphere (constant $\rho \propto r^{-0.4}$) and 22 for the west (high density) hemisphere applying the density gradients as obtained from the DUSTY model. The inner and outer radius were taken from the DUSTY model and were set to 2.08$\times$10$^{14}$~cm and 3.12$\times$10$^{17}$~cm respectively, while the inner H$_{2}$ density was set to 9$\times$$10^{5}$~cm$^{-3}$. This value was derived from the best fit dust model assuming that the gas is entirely molecular and using a mean gas mass per hydrogen of 1.4 amu and a gas--to--dust ratio of 100.

Dust continuum radiation is taken into account using the same OH5 opacities as in the DUSTY model. We assume a static envelope without infall or expansion, a fixed CO abundance of 10$^{-4}$ and a turbulent line width of 3.5~\kms. 


Figure~\ref{fig:ratran} shows the lines of CO 1--0, CO 2--1 and CO 9--8 and isotopologues as observed towards IRS~1 overplotted with the convolved ($\sim$21\as) synthetic emission as calculated with RATRAN. The red color represents the resulting line profiles when using the lower density east hemisphere from the DUSTY models, while the blue color represents the high density west hemisphere. With the exception of the low--$J$ transitions of the isotopologues, we observe no significant differences between the two cases. The fit between observed and modeled lines shows that the low--$J$ lines are reproduced but the higher--$J$ lines are significantly underestimated, especially for the isotopologues. The model cannot reproduce the highly excited lines tracing high column densities of warm and very dense gas. An adapted kinematic structure and/or clumpiness can potentially remove the modeled self--absorption dip of the optically thick $^{12}$CO lines but treating these effects are beyond the scope of this work. Models with T$_{\mathrm gas}$ $>$ T$_{\mathrm dust}$, as inferred from the simple analysis in
Sect. \ref{sect:Comparison_Dust_Gas}, would probably be able to reproduce the high--$J$ lines without
significantly affecting the low--$J$ lines. Local density enhancements
can be an alternative solution, if the total mass of the cloud is
conserved, such as in a clumpy or disk/outflow geometry.


\section{Discussion and conclusions} 


Based on PACS, HIFI and IRAM data, we find a large range of dust and gas temperatures in the S~140 region. The warmest gas ($\sim 55$~K) is around the three infrared sources, while the surrounding environment is characterized by colder temperatures ($\sim$25--35~K). The gas temperature increases towards the southwest edge of the cloud (IF) reaching values of $\sim$40--45~K. This rise of temperature is the result of the UV heating from the external B0V star HD211880 that lies $\sim$7$\arcmin$ southwest of the edge of the cloud. Unfortunately, we cannot compare the gas and the dust temperatures towards IF, because of the smaller area that our continuum dust observations cover. 

We find that the dust density gradient around IRS~1 is likely to be shallower than the
best fits found in some earlier models in order to explain the spatial structure we find in the
far--infrared. The strong emission to the west and southwest of IRS~1 at wavelengths longward of
100 ~$\mu$m can possibly be explained solely by heating from IRS~1 together with a strongly
increasing density to the west and southwest. 
We do, however, find a significantly better fit
by interpreting the emission as arising from two or more separate sources with very different
amounts of dust around the sub--mm source compared to IRS~1. The luminosity of the internal
source at the sub--mm peak is likely to be few $\times$ 10$^{2}$ $\lsun$ in this model. This model with
an internal luminosity source at the sub--mm peak is consistent with the study of \citet{Maud13} who find that the two strongest 1.3 and 2.7 mm emission peaks are the position
of IRS~1 and SMM~1. Since our models did not include outflow cavities or clumpiness, it is possible
that this conclusion should be revisited after more extensive modeling.

The gas temperature analysis including both high and low--$J$ lines was possible only along the HIFI cut and revealed a systematic excess of gas temperatures against dust temperatures. This result indicates a more efficient gas heating even at densities $\geq$ 10$^{4.5}$ cm$^{-3}$ where the gas and dust are usually expected to be in thermal equilibrium. New SOFIA/GREAT mapping observations of CO 13--12 and 16--15 confirm this excess increasing the gap between the two even more (Ossenkopf et al. submitted).
The high--$J$ lines show that a limitation to the low--$J$ analysis (IRAM) causes an underestimate of the gas temperatures by $\sim$7--12~K and thus provides only a lower limit of the temperatures in the cloud. Thus it is most likely that the gas temperature is higher than the dust also in the whole field. Unfortunately, the higher--$J$ lines do not cover the entire region so that we cannot prove this. The detailed gas modeling also indicates that DUSTY results are good estimates of the temperature and density structure around IRS~1 for the colder gas but there is an excess of warm dense gas that cannot be reproduced by these models (i.e. CO 9--8). The attempt to fit those higher transition lines by modeling hotter gas than dust and/or applying different geometries (i.e. clumps, disk geometry) that would increase the density locally but not the total amount of mass is in our future plans. 

The observed gas temperature excess cannot be explained by a higher cosmic ray ionization. \citet{Floris_cr2000} reported a cosmic ray ionization rate of $<$ 10$^{-16}$~s$^{-1}$, which is too low to provide a major gas heating in S~140. Some gas heating may be due to outflow activities. In the RADEX and RATRAN analysis, we focused on quiescent gas and limited the contribution from protostellar outflows. However, it is possible to have high velocity motions perpendicular to the line of sight. In addition, the larger difference between gas and dust appears in points close to the observed outflow by \citet{Preibisch2002}. A collection of shocks, longitudinal and transversal to the walls of the possible inner cavity around the IR sources will contribute to the emission at the ambient velocity. Oblique shock components are thus a plausible scenario for a somewhat more efficient heating of the gas compared to the dust.

The most probable scenario appears to be the deep UV radiation in a clumpy medium. Dust is heated by UV and IR while gas is heated by processes driven 
only by UV with the main one being the photo--electric heating \citep[][]{Hollenbach1997,Roellig2013}. In the vicinity of any
stellar or protostellar source the gas is much hotter than the dust
when dust and gas are not coupled. Dust--gas collisions try to
equilibrate them if the density is high enough. The dust quickly
attenuates the UV so that the gas becomes colder when going away from
IRS~1. The penetration of the IR is deeper (lower extinction in IR
compared to UV), so that the decay of the temperature of the dust is shallower than that of the gas. However, as we observe that the gas remains hotter than the dust over a relatively large distance from IRS~1, the UV seems to penetrate much deeper into the cloud than expected from a homogeneous medium. The natural explanation is a clumpy medium where there are always enough rays between the clumps that still allow UV propagation to keep the gas warm.

Previous studies already report S~140 to be clumpy \citep[e.g.][]{Kramer1998}. 
\citet{Battersby14} performed a similar study in order to determine the relationship between gas and dust in a massive star-forming region (G~32.02$+$0.05) by comparing the physical properties derived from each, using NH$_{3}$ transitions. In that study similar temperature differences have been reported with the dust temperatures being lower than gas temperatures (by a few K) in the quiescent region, indicating that gas and dust might be not well--coupled in such environments. Differences between gas and dust temperatures in several environments with high densities have been reported in more studies \citep[i.e.][]{Papadopoulos2011,Hollenbach1999}.


Our observations trace warm and cold dust but also warm and less warm gas since we have multiple transitions of CO. The low--J lines tend to show the surface of clumps since they are optically thick but the optically thin isotopologues $^{13}$CO 1--0 and C$^{18}$O 1--0 carry information from deeper positions. Furthermore the higher transitions CO 9--8 and $^{13}$CO 10--9 trace warmer gas but they also arise from deeper positions. The fact that we have optically thin lines and the dust emission is optically thin confirms that we trace the whole line of sight both in gas and dust and thus the observational limitations can probably be excluded.


\bibliographystyle{aa} 
\bibliography{ref}

\begin{acknowledgements}
The authors would like to thank the referee for a careful reading of the manuscript and for providing helpful comments and suggestions that improved the paper.
PH was supported through the NASA Herschel Science Center data analysis
funding program,
by a NASA contract issued by the Jet Propulsion Laboratory, California
Institute of Technology to the University of Texas.
VO was supported through the Collaborative Research Center 956,
sub--project C1, funded by the Deutsche Forschungsgemeinschaft (DFG).
AF thanks the Spanish MINECO for funding support from grants CSD2009--00038 and AYA2012--32032.
HIFI has been designed and built by a consortium of institutes and
university departments from across Europe, Canada and the United States under
the leadership of SRON Netherlands Institute for Space Research, Groningen,
The Netherlands and with major contributions from Germany,
France and the US. Consortium members are: Canada: CSA, U.Waterloo; France: CESR, LAB,
LERMA, IRAM; Germany: KOSMA, MPIfR, MPS; Ireland, NUI Maynooth;
Italy: ASI, IFSI--INAF, Osservatorio Astrofisico di Arcetri-INAF; Netherlands:
SRON, TUD; Poland: CAMK, CBK; Spain: Observatorio Astron\'{o}mico Nacional
(IGN), Centro de Astrobiologa (CSIC-INTA). Sweden: Chalmers University
of Technology - MC2, RSS \& GARD; Onsala Space Observatory; Swedish
National Space Board, Stockholm University - Stockholm Observatory;
Switzerland: ETH Zurich, FHNW; USA: Caltech, JPL, NHSC. 
\end{acknowledgements}

\end{document}